\documentclass[10pt]{iopart}
\usepackage{iopams}
\usepackage{subeqn}
\usepackage[numbers, sort&compress]{natbib}

\usepackage{amssymb}
\usepackage{graphicx}
\usepackage{color}
\definecolor{darkgreen}{rgb}{0,0.5,0}
\definecolor{purple}{rgb}{0.5,0,0.5}
\definecolor{blue}{rgb}{0.0,0.0,0.50}
\definecolor{scarlet}{rgb}{1.0,0.2,0}

\usepackage[mathscr,scaled=1.15]{urwchancal}
\DeclareFontFamily{OT1}{pzc}{}
\DeclareFontShape{OT1}{pzc}{m}{it}%
{<-> s * [1.15] pzcmi7t}{}
\DeclareMathAlphabet{\mathpzc}{OT1}{pzc}{m}{it}

\newcommand{\eqref}[1]{(\ref{#1})}

\begin{document}

\topical[The pion: an enigma within the Standard Model]{The pion: an enigma within the Standard Model}

\author{Tanja Horn$^{1,2}$ and Craig D.\ Roberts$^3$}

\address{$^1$ Department of Physics, The Catholic University of America\\ \hspace*{0.9em}Washington DC 20064, USA}
\address{$^2$ Thomas Jefferson National Accelerator Facility\\ \hspace*{0.9em}Newport News, Virginia 23606, USA\\ \hspace*{0.9em}E-mail: hornt@jlab.org}

\address{$^3$ Physics Division, Argonne National Laboratory\\ \hspace*{0.9em}Argonne, Illinois 60439, USA\\ \hspace*{0.9em}E-mail: cdroberts@anl.gov}
\vspace{10pt}
\begin{indented}
\item[]10 February 2016
\end{indented}

\begin{abstract}
Quantum Chromodynamics (QCD) is the strongly interacting part of the Standard Model.  It is supposed to describe all of nuclear physics; and yet, almost fifty years after the discovery of gluons and quarks, we are only just beginning to understand how QCD builds the basic bricks for nuclei: neutrons and protons, and the pions that bind them together.  QCD is characterised by two emergent phenomena: confinement and dynamical chiral symmetry breaking (DCSB).
They have far-reaching consequences, expressed with great force in the character of the pion; and pion properties, in turn, suggest that confinement and DCSB are intimately connected.  Indeed, since the pion is both a Nambu-Goldstone boson and a quark-antiquark bound-state, it holds a unique position in Nature and, consequently, developing an understanding of its properties is critical to revealing some very basic features of the Standard Model.
We describe experimental progress toward meeting this challenge that has been made using electromagnetic probes, highlighting both dramatic improvements in the precision of charged-pion form factor data that have been achieved in the past decade and new results on the neutral-pion transition form factor, both of which challenge existing notions of pion structure.  We also provide a theoretical context for these empirical advances, which begins with an explanation of how DCSB works to guarantee that the pion is unnaturally light; but also, nevertheless, ensures that the pion is the best object to study in order to reveal the mechanisms that generate nearly all the mass of hadrons.
In canvassing advances in these areas, our discussion unifies many aspects of pion structure and interactions, connecting the charged-pion elastic form factor, the neutral-pion transition form factor and the pion's leading-twist parton distribution amplitude.  It also sketches novel ways in which experimental and theoretical studies of the charged-kaon electromagnetic form factor can provide significant contributions.
Importantly, it appears that recent predictions for the large-$Q^2$ behaviour of the charged-pion form factor can be tested by experiments planned at the upgraded 12\,GeV Jefferson Laboratory.  Those experiments will extend precise charged-pion form factor data up to momentum transfers that it now appears may be large enough to serve in validating factorisation theorems in QCD.  If so, they may expose the transition between the nonperturbative and perturbative domains and thereby reach a goal that has driven hadro-particle physics for around thirty-five years.
\end{abstract}

%
\vspace{1pc}
\noindent{\it Keywords}:
Abelian anomaly,
confinement,
dynamical chiral symmetry breaking,
elastic and transition form factors,
$\pi$-meson,
$K$-meson,
non-perturbative QCD,
parton distribution amplitudes and functions
%
%
%
%


\section{Importance and nature of the pion within the Standard Model}
\label{secIntro}
The pion occupies a special place in nuclear and particle physics.  Its existence was predicted in 1935 \cite{Yukawa:1935xg} and after a twelve year search, it became the first meson seen experimentally: the charged pion in 1947 \cite{Lattes:1947mw} and the neutral pion in 1950 \cite{Bjorklund:1950zz, Panofsky:1950gj}.  It is Nature's longest ranging nuclear messenger, being the archetype for meson-exchange forces and, hence, even today plays a critical role as an elementary field in nuclear structure Hamiltonians \cite{Pieper:2001mp, Epelbaum:2008ga, Machleidt:2011zz}.  Yet, after eighty years of study, controversies remain concerning the pion's internal structure, how this connects with newly measured cross-sections and even whether the internal structure of the charged and neutral pion is the same.  The longstanding challenge is to address these and related questions within quantum chromodynamics (QCD) \cite{Marciano:1979wa}, which should describe the strong interaction sector in the Standard Model of Particle Physics.

The trouble with the pion and the source of its great fascination is the apparently dichotomous nature of this peculiar hadron. Since no hadron is elementary, the pion must be a bound-state; and following introduction of the constituent-quark model (CQM) in the early 1960s \cite{GellMann:1964nj, Zweig:1981pd}, the pion came to be considered as an ordinary quantum mechanical bound-state of a constituent-quark and constituent-antiquark.  In that approach, however, explaining its properties requires a finely tuned potential \cite{Godfrey:1985xj}.  In order to see why, consider that CQMs describe a nucleon with mass $m_N = 940\,$MeV as a combination of three constituent-quarks; and a range of successful applications suggest that it is reasonable to infer that each constituent has mass $M_Q \approx M_N/3 \approx 310\,$MeV.  Furthermore, a description of the $\rho$-meson under similar assumptions yields $M_Q = M_{\bar Q} \approx m_\rho/2 \approx  390\,$MeV, \emph{i.e}.\ a consistent outcome.  However, if the pion is added to this mix, one arrives at $M_Q \approx m_\pi/2 \approx  70\,$MeV, \emph{viz}.\ a completely different scale, so that the pion appears unnaturally light.

During the same period, a competing picture of the pion began to emerge via the development of current algebra \cite{GellMann:1964tf, Fubini:1964boa, Adler:1968hc, Weisberger:1966ip} and the notion of partial conservation of the axial current (PCAC) \cite{GellMann:1960np, Nambu:1960xd, Chou:1961, GellMann:1962xb} in order to describe hadronic weak-decays and pion-nucleon interactions.  As the crucial role of the pion in the application of these ideas became appreciated, some practitioners began to ask whether it was natural for the pion to be so light: $m_\pi/m_N \approx 0.15$ came to be seen as an empirical fact in demand of an explanation.  In this connection, it then occurred to some of those involved that there is one special circumstance under which Nature produces unnaturally light spinless bosons, \emph{i.e}.\ when a symmetry is spontaneously or dynamically broken in the underlying theory  \cite{Goldstone:1961eq, Nambu:1960tm, Goldstone:1962es}.  Within the context of current algebra and PCAC, the low-mass pion could then be explained as the (pseudo) Nambu-Goldstone boson that arises as a consequence of the dynamical breakdown of the symmetry associated with the isovector axial-current, \emph{viz}.\ chiral symmetry.  As highlighted by the successes of chiral perturbation theory \cite{Weinberg:1966fm, Weinberg:1967kj, Weinberg:1968de, Coleman:1969sm, Callan:1969sn, Dashen:1969eg, Dashen:1969ez}, this has turned out to be a very useful idea; and dynamical chiral symmetry breaking (DCSB) is now understood to be one of the most important emergent phenomena in the Standard Model, being responsible for the generation of more than 98\% of visible mass \cite{national2012NuclearS, Brodsky:2015aiaS}.

Crucial amongst this train of discoveries and of particular relevance herein is the result established in Ref.\,\cite{GellMann:1968rz}; namely, in a Universe that realises the pion as a pseudo-Nambu-Goldstone boson,
\begin{equation}
\label{oGMOR}
m_\pi^2 \propto \langle \pi | U_{ECSB} | \pi \rangle\,,
\end{equation}
where $U_{ECSB}$ is any term which explicitly breaks chiral symmetry in the Hamiltonian that describes strong interactions.  No picture of the pion founded upon a quantum mechanics of constituent-quarks and -antiquarks can reproduce the Gell-Mann--Oakes--Renner (GMOR) relation, Eq.\,\eqref{oGMOR}.  With careful tuning, such models can produce a light pion; but they always yield $m_\pi \propto \langle \pi | U_{ECSB} | \pi \rangle$.  Notwithstanding that, quantum mechanics models based on the notion of $M_Q \approx M_N/3$ constituent-quarks also achieved numerous successes.  We thus arrive at one of the most significant questions within the Standard Model, \emph{viz}.\ how does one simultaneously realise the GMOR relation and form an almost-massless bound state from very massive constituents?  Stated more generally, how does one expose and express all the consequences of DCSB, both in hadron mass patterns and interactions, within a theory that describes all those hadrons as bound-states of apparently massive quarks (and antiquarks)?

This problem became even more challenging following the formulation of QCD in terms of colour-carrying gluons and quarks, because it was then necessary to reconcile predictions for low-energy processes, such as $\pi \pi$ scattering \cite{Weinberg:1966kf, Colangelo:2001df} and the neutral-pion's two-photon decay \cite{Adler:1969gk, Bell:1969ts}, with perturbative QCD (pQCD) analyses that yield predictions for pion elastic and transition form factors, $F_\pi(Q^2)$ and $G(Q^2)$ respectively, at high energies \cite{Farrar:1979aw, Lepage:1979zb, Efremov:1979qk, Lepage:1980fj}.  The latter can be expressed succinctly.  In the case of the elastic form factor:
\begin{equation}
\label{pionUV}
\exists Q_0>\Lambda_{\rm QCD} \; |\;   Q^2 F_\pi(Q^2) \stackrel{Q^2 > Q_0^2}{\approx} 16 \pi \alpha_s(Q^2)  f_\pi^2 \mathpzc{w}_\varphi^2(Q^2),
\end{equation}
where: $f_\pi=92.2\,$MeV is the pion's leptonic decay constant \cite{Agashe:2014kda}; $\alpha_s(Q^2) $ is the strong running-coupling, which at leading-order is
\begin{equation}
\alpha_s(Q^2) = 4 \pi/[\beta_0\,\ln(Q^2/\Lambda^2_{\rm QCD})],
\label{alphaS}
\end{equation}
with $\beta_0 = 11 - (2/3) n_f$ ($n_f$ is the number of active quark flavours); and
\begin{equation}
\label{wphi}
\mathpzc{w}_\varphi(Q^2) = \frac{1}{3} \int_0^1 dx\, \frac{1}{x} \varphi_\pi(x;Q^2)\,,
\end{equation}
where $\varphi_\pi(x;Q^2)$ is the pion's valence-quark parton distribution amplitude (PDA), which describes the probability that a valence-quark within the pion is carrying a light-front fraction $x$ of the bound-state's total momentum.  Here, $\Lambda_{\rm QCD} \sim 200\,$MeV is the natural mass-scale of QCD (whose dynamical generation through quantisation spoils the conformal invariance of the classical massless theory \cite{Collins:1976yq,Nielsen:1977sy,tarrach}); and, notably, the value of $Q_0$ is not predicted by pQCD.

On the domain $\Lambda_{\rm QCD}^2/Q^2 \simeq 0$  \cite{Lepage:1979zb, Efremov:1979qk, Lepage:1980fj},
\begin{equation}
\label{PDAcl}
\varphi_\pi(x;Q^2) \stackrel{\Lambda_{\rm QCD}^2/Q^2 \simeq 0}{\approx} \varphi_\pi^{\rm cl}(x) = 6 x (1-x)\,;
\end{equation}
and hence
\begin{equation}
\label{Fpicl}
Q^2 F_\pi(Q^2) \stackrel{\Lambda_{\rm QCD}^2/Q^2 \simeq 0}{\approx} 16 \pi \alpha_s(Q^2)  f_\pi^2 \,.
\end{equation}

The result for the transition form factor is even simpler because it does not involve the pion PDA at leading order:
\begin{equation}
\label{BLuv}
\exists \tilde Q_0>\Lambda_{\rm QCD} \; |\;
Q^2 G(Q^2) \stackrel{Q^2 > \tilde Q_0^2}{\approx}  4\pi^2 f_\pi,
\end{equation}
where $\tilde Q_0$ may be distinct from $Q_0$.

The validity of Eqs.\,\eqref{pionUV}, \eqref{BLuv} relies on QCD being a local, relativistic, non-Abelian, quantum gauge-field theory, which possesses the property of asymptotic freedom \cite{Politzer:2005kc, Wilczek:2005az, Gross:2005kv}, \emph{i.e}.\ the QCD interactions are weaker than Coulombic at short distances.  This behaviour is evident in the one-loop expression for the running coupling in Eq.\,\eqref{alphaS} and verified in a host of experiments (see, \emph{e.g}.\ Fig.\,9.4 in Ref.\,\cite{Agashe:2014kda}).  Hence, as a necessary consequence of asymptotic freedom, $\alpha_s(Q^2)$ must increase as $Q^2$ approaches $\Lambda_{\rm QCD}^2$ from above.  In fact, at $Q^2\approx 4\,$GeV$^2=: \zeta_2^2$, which corresponds to a length-scale on the order of 10\% of the proton's radius, it is empirically known that $\alpha_s(\zeta^2)>0.3$.  These observations describe a peculiar circumstance, \emph{viz}.\ an interaction that becomes stronger as the participants try to separate.  It leads one to explore some curious possibilities: If the coupling grows so strongly with separation, then perhaps it is unbounded; and perhaps it would require an infinite amount of energy in order to extract a quark or gluon from the interior of a hadron?  Such thinking has led to the \\[1ex]
\hspace*{4em}\parbox[t]{0.7\textwidth}{\textit{Confinement Hypothesis}: Colour-charged particles cannot be isolated and therefore cannot be directly observed.  They clump together in colour-neutral bound-states.}\\

Confinement seems to be an empirical fact; but a mathematical proof is lacking.  Partly as a consequence, the Clay Mathematics Institute offered a ``Millennium Problem'' prize of \$1-million for a proof that $SU_c(3)$ gauge theory is mathematically well-defined \cite{Jaffe:Clay}, one necessary consequence of which will be an answer to the question of whether or not the confinement conjecture is correct in pure-gauge QCD.  There is a problem with this, however: no reader of this article can be described within pure-gauge QCD.  The presence of quarks is essential to understanding all known visible matter, so a proof of confinement that deals only with pure-gauge QCD is chiefly irrelevant to our Universe.  We exist because Nature has supplied two light quarks and those quarks combine to form the pion, which is unnaturally light and hence very easily produced.  Therefore, no understanding of Standard Model confinement is practically relevant unless the picture also explains the connection between confinement and DCSB, and therefore the existence and role of pions, \emph{i.e}.\ pseudo-Nambu-Goldstone modes with $m_\pi<\Lambda_{\rm QCD}$.

Given the importance of the pion within the Standard Model, enormous experimental effort has been devoted to verifying the entire gamut of rigorous theoretical predictions that relate to pion properties and interactions.  The low-energy results have been tested extensively and hold up well \cite{Batley:2010zza, Larin:2010kq}.  It might be argued, however, that such experiments only check global symmetries and breaking patterns, which could be characteristic of a broad class of theories.  On the other hand, as emphasised by the quark discovery experiments performed at the Stanford Linear Accelerator Center (SLAC) \cite{Taylor:1991ew, Kendall:1991np, Friedman:1991nq}, measurements at high-energy are a direct probe of QCD itself.  In large part, this notion is behind the construction and operation of the Thomas Jefferson National Accelerator Facility (JLab), in Newport News, Virginia, USA, the relativistic heavy-ion collider (RHIC) on Long Island, New York, USA, and also the rationale for numerous measurements of hard exclusive processes in the BaBar experiment at SLAC and with the Belle detector at the high energy accelerator research organisation (KEK) in Tsukuba, Japan.  We will focus herein on some of the tensions between experiment and theory found in connection with such measurements; and describe recent theoretical advances that may relieve them.


\section{Form factors: experimental history, current status, and future}
\label{sec:pion-ff-exp}
Form factors are of primary importance in hadron physics because they provide Poincar\'e-invariant information about the nonpointlike nature of QCD's observable bound-states, and the distribution of gluons and quarks within them.  At low values of $Q^2$, the pion's elastic form factor,  $F_{\pi}$, has been determined directly up to photon energies of $Q^2 =0.253\,$GeV$^2$ at Fermilab \cite{Dally:1982zk, Dally:1981ur} and at the CERN SPS \cite{Amendolia:1986wj, Amendolia:1984nz} from the scattering of high-energy, charged pions by atomic electrons.  These data were used to constrain the charge radius of the pion, which is determined to be $r_\pi=0.657 \pm 0.012\,$fm.  Owing to kinematic limitation in the energy of the pion beam and unfavorable momentum transfer, one has to resort to other experimental methods to reach the higher $Q^2$ regime.  At higher values of $Q^2$, $F_{\pi}$ can be determined from electroproduction of pions on the proton.

In general, electroproduction reactions are of interest as they allow for measuring photoproduction amplitudes as functions of the photon mass.  The weakness of the electromagnetic interaction allows one to treat these reactions in the one-photon exchange approximation as virtual photoproduction by space-like photons, $Q^2>$0, whose mass, energy, direction, and polarization density are tagged by the scattered electron \cite{Hand:1963bb}.  The electroproduction reaction can be described in terms of form factors, which are generalizations of the form factors observed in elastic electron-hadron scattering or in terms of cross-sections that are extensions of the photoproduction cross-sections. In general, the virtual photon is polarised.  There are two transverse polarization states and a third component, which can either be taken as a scalar or as longitudinal, with its only component along the momentum-direction of the virtual photon.

For a coincidence experiment in which the scattered electron and the electroproduced charged pion are detected, the differential cross-section can be expressed in terms of a known electrodynamic factor and a virtual photoproduction cross section.  The latter can be expressed in terms of linear combinations of the products of virtual-photoproduction helicity amplitudes, which are the unpolarised transverse production, the purely scalar (longitudinal) production, and the interference terms between the transverse and transverse-scalar states.  This reduced cross-section can be written as a sum of four separate cross-sections or structure functions, which depend on $Q^2$, and also the invariant mass of the virtual photon-nucleon system,$W$, and the Mandelstam variable, $t$:
\begin{eqnarray}
\label{eq:sepsig}
  2\pi \frac{d^2 \sigma}{dt d\phi} & = & \frac{d \sigma_T}{dt} + \epsilon  \frac{d \sigma_L}{dt} \\ \nonumber
                                   & + & \sqrt{2 \epsilon (1 + \epsilon)}  \frac{d \sigma_{LT}}{dt} cos \phi
                                    +  \epsilon  \frac{d \sigma_{TT}}{dt} cos 2 \phi \, .
\end{eqnarray}
Here, $\epsilon=1/\left(1+\frac{2 |\mathbf{q}|^2}{Q^2} \tan^2\frac{\theta_{e}}{2}\right)$ is the polarization of the virtual photon, with $\mathbf{q}$ denoting the three-momentum of the transferred virtual photon, $\theta_{e}$ is the electron scattering angle, and $\phi$ is the angle between the scattering plane defined by the incoming and scattered electrons and the reaction plane defined by the transferred virtual photon and the scattered meson.  In order to separate the different structure functions one has to determine the cross-section for at least two sufficiently different values of $\epsilon$ as a function of the angle $\phi$ for fixed values of $W$ and $t$.

The determination of the pion form factor from pion electroproduction requires that one-meson exchange (the pion pole) dominates the longitudinal cross-section at small values of $t$.  The pion pole includes a factor $-t$/$(t-m^2_\pi)^2$, which is zero at $t$=0 and reaches a maximum at $t$=$-m^2_\pi$.  The first value is unphysical, since forward scattering occurs at $t_{min}$=$-4m_p^2 \xi^2/(1-\xi^2)$, where $m_p$ is the proton mass and $\xi=x_B/(2-x_B)$, with $x_B=\,$Bjorken-$x$, while the second can be reached in experiments for $\xi \sim m_{\pi}/2m_p$.  The dominance of the pion pole in the longitudinal cross-section and its characteristic $t$--dependence allows for extractions of the electromagnetic pion form factor from these data.

Cross-section data suggest a dominant pion pole in the longitudinal $\pi^+$ cross-section at values of $-t<0.3\,$(GeV/c)$^2$ \cite{Favart:2015umi}.  The strength of the pion pole falls-off rapidly with increasing values of $t$.  The observation of a dominant pion pole alone is, however, insufficient to make a precise extraction of $F_{\pi}$ from the data.  In order to minimise background contributions, the longitudinal cross-section, $\sigma_L$, is thus isolated via a Rosenbluth L/T(/LT/TT) separation.  Without an explicit L/T separation it is not clear what fraction of the cross-section is due to longitudinal photons and what part owes to the contribution of the pole in these kinematics.  Data from survey experiments, like those from Refs.\,\cite{Airapetian:2007aa, Park:2012rn}, though interesting in their own right, are thus not used in precision form factor extractions.

Data for $F_{\pi}$ using the pion electroproduction reaction have been obtained for values of $Q^2$ up to 10\,GeV$^2$ at Cornell \cite{Bebek:1974ww, Bebek:1976qm, Bebek:1977pe}.  However, those data suffer from relatively large statistical and systematic uncertainties.  More precise data were obtained at the Deutsches Elektronen-Synchrotron (DESY) \cite{Ackermann:1977rp, Brauel:1979zk}.  With the availability of high-intensity electron beams, combined with accurate magnetic spectrometers at JLab, it has been possible to determine L/T-separated cross-sections with high precision.  The measurement of these cross-sections in the regime of $Q^2=0.60-1.60\,$GeV$^2$ (Experiment Fpi-1 \cite{Volmer:2000ek, Tadevosyan:2007yd}) and $Q^2=1.60-2.45\,$GeV$^2$ (Experiment Fpi-2 \cite{Horn:2006tm} and pionCT \cite{Horn:2007ug}) are described in detail in Ref.\,\cite{Blok:2008jy}.  Pion electroproduction experiments are performed at the smallest possible value of $t$, which is still a distance away from the pion pole, and thus the pion form factor has to be obtained either by kinematic extrapolation or by using a theoretical model.  In the latter case, consistency between data and model is essential.

Frazer \cite{Frazer:1959zz} originally proposed that the pion form factor could be extracted from $\sigma_L$ via a kinematic extrapolation to the pion pole, and that this could be done in an analytical manner using the ``Chew-Low extrapolation'' \cite{Chew:1958wd}.  The last serious attempt to extract $F_{\pi}$ from electroproduction data using this method is described in Ref.\,\cite{Devenish:1973ta}.  The extrapolation failed to produce a reliable result, because different polynomial fits that were equally likely in the physical region gave divergent values of the form factor when extrapolated to the pion pole.

Brown \emph{et al}., at the Cambridge electron accelerator, were the first to use a theoretical model to extract $F_{\pi}$ from electroproduction data.  In Ref.\,\cite{Brown:1973wr} the model of Berends \cite{Berends:1970ti} was used.  This model includes the dominant isovector Born term, with corrections for $t$ values away from the pole by means of fixed-$t$ dispersion relations, and gave a fair description of the data.  However, the LT term of the cross-section and the $t$-dependence of the data were systematically underpredicted.  The DESY experiments \cite{Ackermann:1977rp, Brauel:1979zk} produced high-quality L/T-separated electroproduction cross-sections and used the generalised Born Term Model (BTM) of Gutbrod and Kramer \cite{Gutbrod:1973qr}, which gave a better overall description of the data, in order to extract $F_{\pi}$ .

More recent experiments like those discussed in Refs.\,\cite{Horn:2006tm, Horn:2007ug, Huber:2008id, Blok:2008jy, Guidal:1997hy} have used a Regge model for pion electroproduction developed by Vanderhaeghen, Guidal and Laget (VGL) \cite{Vanderhaeghen:1997ts, Guidal:1997by, Guidal:1997hy} for the extraction of $F_{\pi}$.  In this model the pole-like propagators of Born term models are replaced with Regge propagators, \emph{i.e}.\ the interaction is effectively described by the exchange of a family of particles with the same quantum numbers instead of a single particle.  The model parameters are fixed from pion photoproduction data.  The free parameters are $\Lambda_{\pi}$ and $\Lambda_{\rho}$, the trajectory cutoffs.  A fit to the model longitudinal cross-section then gives $F_\pi$ at each value of $Q^2$.  If the same vertices and coupling constants are used, the Regge model and the BTM calculations agree at the pole of the exchanged particle; but away from the pole, where data are acquired, the Regge model provides a superior description of the available data.

\begin{figure}[t]
\centerline{\includegraphics[clip,width=0.6\textwidth]{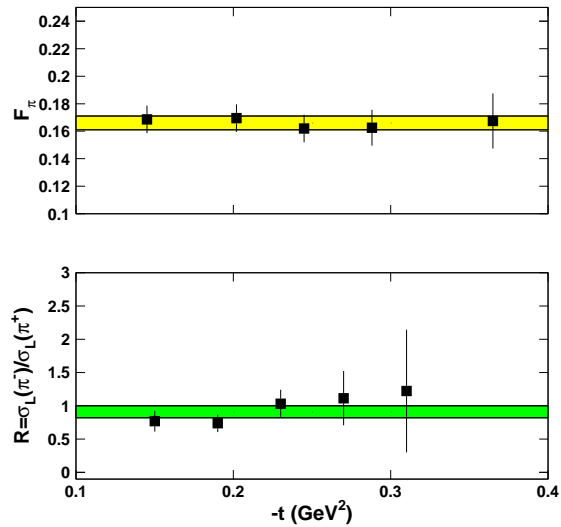}}
\caption{\label{Fig0t}
\textbf{Upper panel} -- Checking $t$- and model-dependence in the pion form factor extraction at a central value of $Q^2=2.45\,$GeV$^2$ and center-of-mass energy $W=2.22 \,$GeV: the solid squares denote the $F_{\pi}$ values for the case in which the model was fit to each point separately and the band shows the $F_{\pi}$ value obtained from a fit to all points.  The error bars and the error band include statistical and uncorrelated uncertainties.  (Data from Ref.\,\cite{Huber:2008id}.)
\textbf{Lower panel} -- Checking dominance of the $t$-channel process in $\sigma_L$ through the charged-pion longitudinal cross-section ratios at $Q^2=2.45\,$GeV$^2$ and $W=2.22 \,$GeV .  The cross-section ratios are close to unity and much larger than the ratios typically found for the transverse cross-section, which are close to $1/4$.  This significant difference suggests pion pole dominance in the longitudinal cross sections (and parton model dominance in the transverse).  The error bars include statistical and uncorrelated uncertainties, and the (green) band denotes the uncertainty of a constant fit to all data points.  (Data from Ref.\,\cite{Huber:2014ius}).
}
\end{figure}

To give confidence in the electroproduction method yielding the physical form factor, one can carry out several experimental studies.  These include checking the consistency of the model with the data, extracting the form factor at several values of $t_{min}$ for fixed $Q^2$, verifying that the pole diagram is the dominant contribution to the reaction mechanism, and verifying that the electroproduction technique gives results consistent with those from $\pi-e$ elastic scattering at the same $Q^2$.   In Ref.\,\cite{Huber:2008id}, to check if the VGL Regge model properly accounts for the pion production mechanism, spectator nucleon, and other off-shell ($t$-dependent) effects, $F_{\pi}$ was extracted at different distances in $t$ from the pion pole.  The resulting $F_{\pi}$ values agreed to within 4\% and did not depend on the $t$ acceptance, which lends confidence in the applicability of the VGL model to the kinematic regime of the data and the validity of the extracted $F_{\pi}$ values.  The dominance of the $t$-channel process in $\sigma_L$ was verified in Ref.\,\cite{Huber:2014ius} through the charged-pion longitudinal cross-section ratios, $R_L = \sigma_L[n(e,e^\prime \pi^-)p]/\sigma_L[p(e,e^\prime \pi^+)n]$, obtained with a deuterium target.  The $t$-channel diagram is purely isovector and so any isoscalar background contributions, like $b_1$(1235) to the $t$ channel, will dilute the ratio.  With increasing $t$, the transverse cross-section ratio is expected to approach the ratio of quark charges, \emph{i.e}.\ 1/4.  The data show that $R_L$ approaches the pion charge ratio, consistent with pion-pole dominance.  The $t$-dependence of the transverse cross section ratio, $R_T=\sigma_T[n(e,e^\prime \pi^-)p]/\sigma_T[p(e,e^\prime \pi^+)n]$, shows a rapid fall-off consistent with $s$-channel quark knockout.  
Illustrations of $t$-channel dominance, and $t$- and model-dependence in the $F_{\pi}$ extraction for $t\in[0.15,0.37]\,$GeV$^2$ are shown in Fig.\,\ref{Fig0t}.
Direct comparison of $F_{\pi}$ values extracted from very low $t$ electroproduction with the exact values measured in elastic $\pi-e$ scattering showed that the data are consistent within the uncertainties, lending further confidence to the validity of the extracted $F_{\pi}$ values.

\begin{figure}[t]
\rightline{\begin{minipage}[t]{0.8\textwidth}
\begin{minipage}{0.75\textwidth}
\centerline{\includegraphics[clip,width=0.97\textwidth]{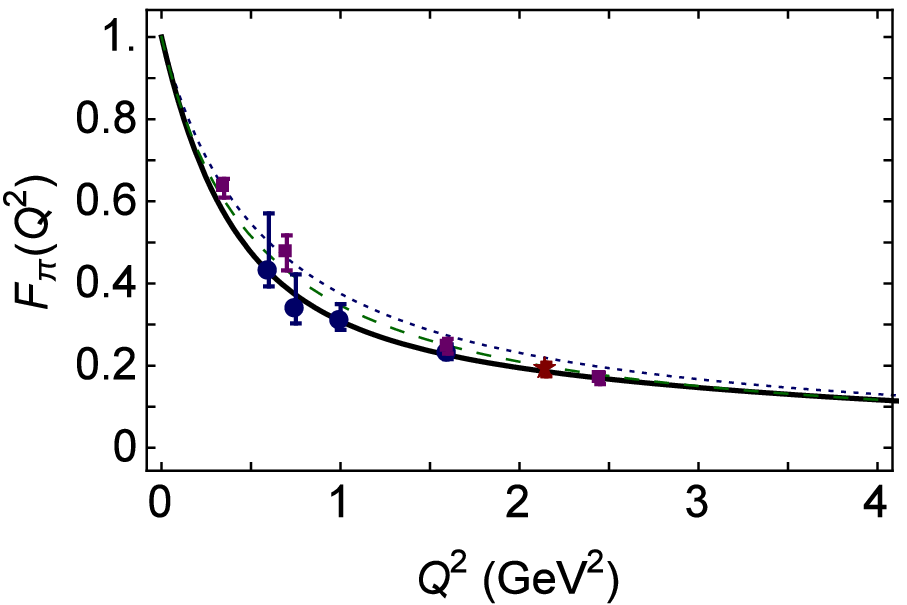}}
\end{minipage}
\begin{minipage}{0.75\textwidth}
\centerline{\includegraphics[clip,width=0.97\textwidth]{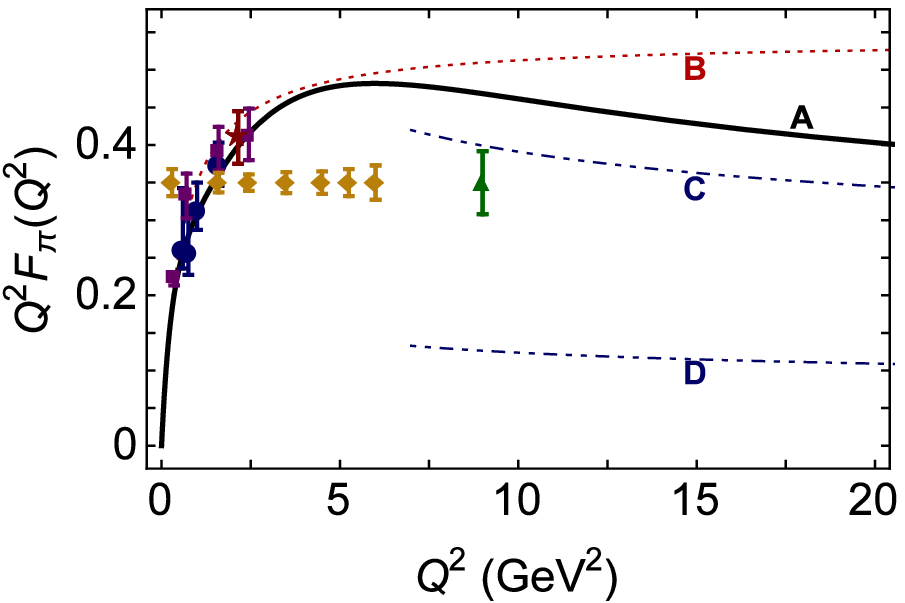}}
\end{minipage}
\end{minipage}}
\caption{\label{Fig1CLpion}
\textbf{Upper panel}.
\emph{Solid curve} -- Charged pion form factor computed in Ref.\,\cite{Chang:2013niaS} ($r_\pi = 0.66\,$fm \emph{cf}.\ experiment \cite{Agashe:2014kda} $r_\pi = 0.672 \pm 0.008\,$fm); \emph{long-dashed curve} -- calculation in Ref.\,\protect\cite{Maris:2000sk}; and \emph{dotted curve} -- monopole form ``$1/(1+Q^2/m_\rho^2)$,'' where $m_\rho=0.775\,$GeV is the $\rho$-meson mass.  In both panels, the filled-star is the point from Ref.\,\cite{Horn:2007ug}, and the filled-circles and -squares are the data described in Ref.\,\protect\cite{Huber:2008id}.
\textbf{Lower panel}.
$Q^2 F_\pi(Q^2)$.  \emph{Solid curve}\,(A) -- prediction in Ref.\,\cite{Chang:2013niaS}.
Remaining curves, from top to bottom: \emph{dotted curve}\,(B) -- monopole form fitted to data in Ref.\,\protect\cite{Amendolia:1986wj}, with mass-scale $0.74\,$GeV;
\emph{dot-dot--dashed curve}\,(C) -- pQCD prediction obtained from Eq.\,\protect\eqref{pionUV} using the modern, dilated pion PDA in Eq.\,\protect\eqref{phiRLreal};
and \emph{dot-dot--dashed curve}\,(D) --  pQCD prediction computed with the conformal-limit PDA in Eq.\,\eqref{PDAcl}, which had previously been used to guide expectations for the
asymptotic behaviour of $Q^2 F_\pi(Q^2)$.
The filled diamonds and triangle indicate the projected reach and accuracy of forthcoming experiments  \protect\cite{E1206101, E1207105}.
}
\end{figure}



The world precision $F_{\pi}$ data are summarised in Fig.\,\ref{Fig1CLpion}.  They seem to follow a monopole form (curve-B, lower panel), which is the maximum allowed value of the form factor as set by the long-distance-scale pion charge-radius.  As explained in Sec.\,\ref{secIntro}, there is a perspective from which one might view the pion bound-state as a quantum mechanical two-body problem and hence there have been numerous computations of $F_\pi(Q^2)$ over many years.  We provide a contemporary update in Sec.\,\ref{SecElasticPion}.  Data on $F_{\pi}$ are also available at timelike momenta, e.g.\ Refs.\,\cite{Pedlar:2005sj, Seth:2012nnS}, out to $t\approx18\,$GeV$^2$.  Here, the form factor is measured via annihilation $e^+ e^- \rightarrow h^+ h^-$, where $h^+ h^-$ is $\pi^+ \pi^-$.  In the analysis of such data, contributions  to the $\pi^+ \pi^-$ sample from the leptonic background and the $\phi(2S)$ resonance must be taken into account.

Just as in the case of elastic scattering, electroproduction and timelike data enable extraction of the pion charge radius, \emph{e.g}.\ Refs.\,\cite{Miller:2009qu, Carmignotto:2014rqa} (spacelike) and Refs.\,\cite{Miller:2010tz, Miller:2011du} (timelike).  In the latter, extension to the spacelike domain is accomplished through the use of dispersion relations and models to obtain separate real and imaginary parts.  In Ref.\,\cite{Carmignotto:2014rqa}, the use of models is avoided and the impact of new experiments is evaluated.  Naturally, in a non-relativistic model the hadron charge distribution is the Fourier transform of the form factor and the proton's distribution was extracted in this manner in Ref.\,\cite{Crawford:2010gv}.  However, in particular for the low-mass pion, this non-relativistic Fourier transform interpretation may be questioned \cite{Miller:2010nz}.  An alternative interpretation has been developed in terms of the transverse charge distribution, and both the proton and pion transverse charge densities have been extracted using this method \cite{Venkat:2010by, Carmignotto:2014rqa}.  The first attempt at an analysis with spacelike data suggests that the transverse charge density of the pion is greater than that of the proton close to the core and that the two densities coalesce for impact parameters larger than 0.3\,fm.  The former is expected because the pion radius is smaller than that of the proton, whereas the latter may be interpreted as the proton consisting of a non-chiral core occupying most of the volume and a meson cloud at large impact parameters.

The pion transition form factor provides the simplest structure for experimental perturbative QCD analysis.  In the experiment one measures the $e^+ e^- \rightarrow e^+ e^- \pi^0$ reaction, where $\gamma^* \gamma \rightarrow \pi^0$ is purely a quantum electrodynamics process.  The detected lepton scatters at large angles yielding a virtual photon with large $Q^2$.  The other lepton, which is not detected, scatters at small angles yielding a nearly real photon.  Data from BaBar \cite{Aubert:2009mc} on the pion transition form factor showed a continuous rise above the asymptotic limit on  $Q^2 \gtrsim10\,$GeV$^2$, which does not conform with the standard QCD approach based on collinear factorization.  This deepened the mystery surrounding the way that QCD makes the changeover from the perturbative to the nonperturbative regime.  More recent measurements from Belle are consistent with QCD scaling and do not show a large $Q^2$ enhancement above $Q^2 \sim 10\,$GeV$^2$.  These data are in agreement with previous data from CELLO/CLEO \cite{Behrend:1990sr, Gronberg:1997fj} and are fully consistent with the $\eta$, $\eta^\prime$ transition form factors \cite{Uehara:2012ag}.  The results from Belle also agree with those from BaBar in the region $Q^2< 9\,$GeV$^2$ \cite{Balakireva:2011wp}.  A statistical analysis of both data sets showed that one cannot predict the trends observed at Belle and BaBar from the other \cite{Stefanis:2012yw}.  Additional data on transition form factors and other exclusive processes are required to reconcile the opposing tendencies observed in the data.  We canvass these and related theoretical issues in Sec.\,\ref{SecTransition}.

\begin{figure}[t]
\rightline{\begin{minipage}[t]{0.8\textwidth}
\begin{minipage}{0.75\textwidth}
\centerline{\includegraphics[clip,width=0.97\textwidth]{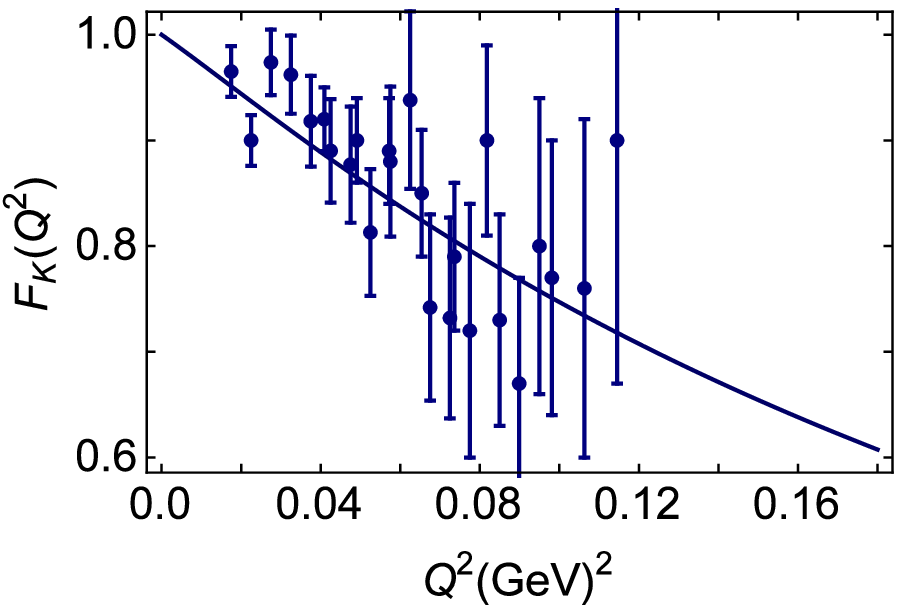}}
\end{minipage}
\begin{minipage}{0.75\textwidth}
\centerline{\includegraphics[clip,width=0.97\textwidth]{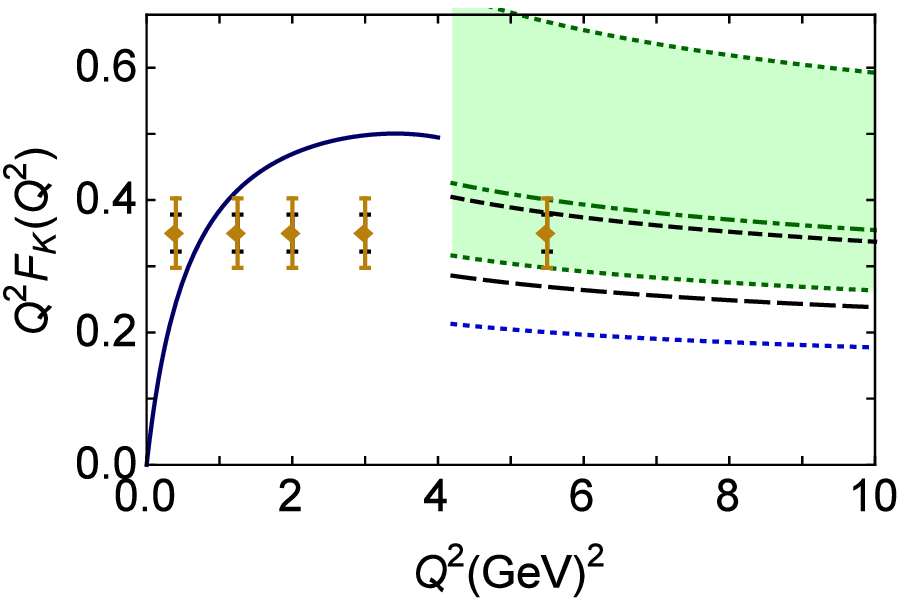}}
\end{minipage}
\end{minipage}}
\caption{\label{fig:fk_exp_current}
\textbf{Upper panel}.
Available data on $F_K(Q^2)$, the kaon elastic electromagnetic form factor \cite{Dally:1980dj, Amendolia:1986ui}.  \emph{Solid (blue) curve} -- Charged kaon form factor computed in Ref.\,\cite{Maris:2000sk}.
\textbf{Lower panel}.
$Q^2 F_K(Q^2)$.  \emph{Solid (blue) curve} -- prediction in Ref.\,\cite{Maris:2000sk}, which does not extend beyond $Q^2=4\,$GeV$^2$ owing to weaknesses in the numerical method.
The remaining curves depict results obtained from the hard-scattering formula in Eq.\,\eqref{kaonUV} when different kaon PDAs are used:
\emph{long-dashed (black) curve} -- DSE prediction, Eq.\,\eqref{DSEkaon};
\emph{dot-dashed (green) curve} -- Eq.\,\eqref{lQCDkaon}, inferred from lQCD values of the lowest two moments, with the green band indicating the uncertainty in the prediction for $Q^2 F_K(Q^2)$ owing to the errors on these moments;
\emph{dashed (black) curve} -- Eq.\,\eqref{PDAKDSEmod}, obtained from the long-dashed curve by supposing that the second moment of the PDA is just 10\% larger;
and \emph{dotted (blue) curve} -- Eq.\,\eqref{PDAcl}, the conformal-limit PDA.
The filled diamonds indicate the projected reach and accuracy of data on $Q^2 F_K(Q^2)$ that are anticipated from a forthcoming experiment \cite{E12-09-011}: two error estimates are pictured, based on different assumptions about the $t$- and model-dependence of the form factor extractions, with the larger uncertainty being conservative.
}
\end{figure}

The next simplest meson available for experimental studies is the kaon, which also contains strangeness.  Similar to pions, the kaon form factor has been determined directly up to photon energies of $Q^2=0.10\,$GeV$^2$ at Fermilab \cite{Dally:1980dj} and at the CERN SPS \cite{Amendolia:1986ui} from the scattering of high-energy, charged kaons by atomic electrons.  These data were used to constrain the mean-square charge radius of the kaon, which is determined to be $r_K^2=0.34\pm 0.05\,$fm$^2$.  The kaon charge radius is typically predicted to be smaller than that of the pion owing to the presence of the heavier strange quark \cite{Tarrach:1979ta, Ji:1990rd, Maris:2000sk, Chen:2012txaS, Ninomiya:2014kja}.  The available $F_{K}$ data are summarised in Fig.\,\ref{fig:fk_exp_current}.

At higher energies the kaon form factor can, in principle, be extracted from kaon electroproduction data.  However, there are experimental challenges that have to be addressed.  In particular, the kaon pole is farther from the physical region than the pion, something which may raise doubts about the ability to obtain reliable information on $F_K$ from kaon electroproduction data.  In the face of such doubts, we note two things.  First, extraction of the pion form factor from similar pion electroproduction data at small $t$ is completed by carefully studying the model dependence of the analysis, not by direct extrapolation; and this justifies greater confidence in this method.  Second, initial comparative extractions of the pion form factor from low-$t$ and larger-$t$ data suggest only a modest model dependence; and the larger-$t$ pion data lie at a similar distance from the pole as most of the projected kaon data discussed herein.  It is thus reasonable to imagine that $F_K$ may be extracted from kaon electroproduction data in a similar fashion, albeit with a larger model dependence than $F_\pi$.

Notwithstanding these observations, detailed demonstration of the dominance of the kaon pole is required if one is to be confident in the extraction of $F_K$.  Current electroproduction data show that the $t$-dependence of the longitudinal kaon cross section is less steep than that of the pion \cite{Horn:2012zza, Favart:2015umi}.  These data include L/T-separated cross sections up to photon energies of $Q^2=2.35\,$GeV$^2$ from Refs.\,\cite{Mohring:2002tr, Coman:2009jk}.  Additional kaon electroproduction data from Jefferson Lab are being analysed.  The results from Refs.\,\cite{Horn:2012zza, Favart:2015umi} are consistent with the kaon pole factor, $-t/(t-m^2_K)^2$, giving less enhancement than that of the pion.  However, calculations predict a small maximum in the kaon cross section near $t$=0.1 GeV$^2$, owing to the kaon pole \cite{Goloskokov:2009ia, Goloskokov:2011rd}.  This value of $t$ is smaller than what has hitherto been possible to reach with current experimental data owing to the lack of suitable experimental facilities.  Access to small values of $t$ and L/T separations of the kaon cross-section will be possible at the JLab\,12 facility  \cite{E12-09-011}.

Current experiments have established the techniques for meson electroproduction experiments and the determination of the meson form factors from these data. The 12\,GeV upgrade at Jefferson Lab (JLab\,12) features new instrumentation that allows for pushing precision meson form factor measurements to the highest momentum transfers to date.  Planned experiments aim for precision measurements of the pion form factor to $Q^2=6\,$GeV$^2$ and also have the potential to determine the pion form factor up to $Q^2\sim 9\,$GeV$^2$ \cite{E1206101, E1207105}. These measurements are made possible by the combination of the two moderate acceptance, magnetic spectrometers in Hall C. The ``High Momentum Spectrometer'' (HMS) provides angular acceptance of 6 msr and can detect particles with momenta up to 7\,GeV/c.  The new ``Super-High Momentum Spectrometer'' (SHMS) features a solid angle of about 4 msr, a momentum coverage up to 11\,GeV/c, and covers scattering angles between 5.5 and 40 degrees.
The small scattering angle capability, combined with excellent control of systematic uncertainties, kinematic reproducibility and well-understood acceptance, enables precision measurements of cross-sections and L/T separations at high luminosity ($>10^{38}$/cm$^2$s).  Such data will provide access to the pion form factor on a domain of momenta that is nearly four-times larger than that explored hitherto.  QCD backgrounds in these data at high values of $Q^2$ and $t$, such as those described in Ref.\,\cite{Carlson:1990zn}, can be isolated experimentally through measurements of the charged-pion ratio, as discussed above, or of the neutral-pion cross section.  For example, since the charged-pion $t$-channel diagram is purely isovector, contamination by isoscalar backgrounds would be visible in a distortion of the ratio $R_L$.  Complementing this, one can capitalise on the fact that the neutral pion production reaction does not have a large pion pole contribution.  Therefore, measurements of the longitudinal cross-section for exclusive neutral-pion production, like those in Ref.\,\cite{E12-13-010}, should be sensitive to non-pole contributions in the charged-pion cross section.

The small-angle capability of the magnetic spectrometers and availability of electron beam energies of 11\,GeV will also allow for measurements of the L/T-separated kaon cross-section at the highest values of $Q^2$ to date \cite{E12-09-011}.  The challenge in extracting the kaon form factor from cross-section data is that one has to demonstrate that the kaon pole dominates.  This can be done in a similar way as for the pion.  In addition, the dominance of the kaon pole can, in principle, be tested through the ratio of the longitudinal cross-sections of $\Sigma^0$ to $\Lambda$ channels.  The SHMS and HMS missing mass resolution is expected to be very good and together with the spectrometer coincidence acceptance will allow for simultaneous measurements of the $\Lambda$ and $\Sigma^0$ channels.  If the kaon pole dominates, the ratio should be similar to the following ratio of coupling constants: $g^2_{pK\Lambda}$/$g^2_{pK\Sigma}$.

The relative contribution of longitudinal and transverse terms to the meson production cross-section and their $t$ and $Q^2$ dependence are also of interest in evaluating the potential for probing a nucleon's transverse spatial structure through such processes.  In general, only when experimental evidence for the onset of leading-twist behaviour is established can one be confident in using the light-front handbag formalism discussed in Refs.\,\cite{Ji:1996nm, Radyushkin:1996nd, Radyushkin:1996ru}.  One of the most stringent experimental tests is the $Q^2$ dependence of the longitudinal meson cross-section.  In the regime where the leading-twist formalism is applicable $\sigma_L$ is predicted to scale as $1/Q^{6}$, the transverse cross-section is expected to scale as $\sigma_T \sim 1/Q^{8}$ and consequently, $\sigma_L >> \sigma_T$ \cite{Collins:1996fb}.

The leading-twist, lowest-order calculation of the $\pi^+$ longitudinal cross-section underestimates current data by an order-of-magnitude \cite{Favart:2015umi}.  This implies that the data are not in the regime where a leading-twist analysis applies.  However, measurements of the fully-separated meson cross-sections and their $(Q^2,t)$-dependence are fundamental and important in their own right.  Fully separated cross-sections are essential for understanding dynamical effects in both variables and in the interpretation of non-perturbative contributions at experimentally accessible kinematics.  Such measurements of L/T-separated cross-sections will be enabled by JLab\,12, extending the current kinematic reach of $\pi^+$ data and including additional systems \cite{E12-07-105, E12-09-011, E12-13-010}.  These data will play an important role in developing our understanding of meson pole dominance and meson form factor extractions, and may also provide experimental evidence that supports interpretation of the data using the light-front handbag formalism.

Recent pion cross-section data \cite{Horn:2006tm, Horn:2007ug, Airapetian:2009ac, Bedlinskiy:2012be, Bedlinskiy:2014tvi, Collaboration:2010kna} suggest that transversely polarised photons play an important role in charged and neutral pion electroproduction.  L/T-separated $\pi^+$ data show a large $\sigma_T$ even at values of $Q^2=2.5\,$GeV$^2$ and $t<0.3\,$(GeV/c)$^2$.  In the HERMES experiment at DESY, a large $\sin (\phi-\phi_S)$  modulation was observed in the Fourier amplitude or transverse target spin asymmetry, $A_{UT}(\sin (\phi-\phi_s))$, which does not seem to vanish in the forward direction \cite{Airapetian:2009ac}.  The observed behaviour of the $A_{UT}$ data demands a strong contribution from transverse photons.  The transverse-transverse interference term in $\pi^0$ production is large in absolute value, suggesting that transverse photons play an important role in this kinematic regime.

In order to interpret the data including a large contribution from transverse photons, the light-front handbag approach has been extended \cite{Goloskokov:2009ia,Goloskokov:2011rd} to transition amplitudes represented by convolutions of transversity generalised parton distributions (GPDs) and subprocess amplitudes calculated with a twist-3 pion wave function.  With such a parametrization in terms of transversity GPDs, both the trends and magnitudes of the $\pi^+$ data and the interference terms of the $\pi^0$ data from JLab and HERMES are described well.  Transversity GPDs in pion electroproduction have also been discussed in \cite{Ahmad:2008hp, Goldstein:2012az, Goldstein:2013gra}.

To confirm the estimates of the contribution of transverse photons and the potential to access GPDs in meson production requires a full separation of the cross-section.  The trends discussed above depend on both $Q^2$ and $t$, and thus it is important to cover as large a kinematic range as possible, including the regime $t<0.3\,$(GeV/c)$^2$.  The first L/T-separated $\pi^0$ cross sections were measured Hall A at JLab-6\,GeV and are under analysis \cite{E07-007}.  These data cover a range in $Q^2$ between 1.5 and 2 GeV$^2$ and $x$ of 0.36.  A larger kinematic coverage for both charged and neutral pion (and kaon) production can be achieved with approved experiments at JLab\,12 \cite{E12-07-105, E12-09-011, E12-13-010}. If experimental evidence for the dominance of $\sigma_T$ can be demonstrated to hold, one may interpret these data using the light-front handbag formalism.

The new experiments described herein are part of an extensive and diverse range of hadron structure experiments planned at JLab\,12 \cite{Dudek:2012vr, Aznauryan:2012baS}.  In the context of this discussion, we would also like to note that experimental data on pion and kaon parton distribution functions is sparse.  It has only been obtained in mesonic Drell-Yan scattering from nucleons in heavy nuclei, with information on the pion's PDFs reported in Refs.\,\cite{Badier:1983mj, Betev:1985pg, Conway:1989fs} and results for the ratio of kaon and pion distribution functions presented in Ref.\,\cite{Badier:1980jq}.  Newer data is not available; but would be welcome, owing to persistent doubts about the large Bjorken-$x$ behaviour of the pion's valence-quark PDF \cite{Holt:2010vj} and because a single modest-quality measurement of the kaon-to-pion ratio cannot be considered definitive.  An approved experiment \cite{Keppel:2015, McKenney:2015xis}, using tagged deep inelastic scattering at JLab\,12 should contribute to a resolution of the pion question; and a similar technique might also serve for the kaon.  Furthermore, new mesonic Drell-Yan measurements at modern facilities could yield valuable information on $\pi$ and $K$ PDFs \cite{Londergan:1995wp, Londergan:1996vh}, as could two-jet experiments at the large hadron collider \cite{Petrov:2011pg}; and, looking further ahead, an electron ion collider (EIC) would be capable of providing access to pion and kaon structure functions through measurements of forward nucleon structure functions \cite{Holt:2000cv, Accardi:2012qutSd}.  A contemporary theoretical discussion of the importance of such measurements is presented in Ref.\,\cite{Chen:2016sno}.

\section{Nambu-Goldstone Mode}
\label{secTheoryA}
Let us return now to Eq.\,\eqref{oGMOR} and address the issue of how this is realised in the Standard Model.  The basic relation here is the axial-vector Ward-Green-Takahashi identity (AVWGTI) \cite{Ward:1950xp, Green:1953te, Takahashi:1957xn, Takahashi:1985yz}, which expresses the nature of chiral symmetry in QCD and describes the pattern by which it is broken.  Of course, this statement is meaningless unless one first understands chiral symmetry.  Textbooks provide useful background at this point, but let us augment that with the following statement: any theory with fermions, which possesses a well-defined chiral limit, can be separated into two distinct theories, one with only left-handed fermions and the other with only right-handed fermions; and no interaction in the original Lagrangian can induce communication between these fermions of different chirality.

QCD is asymptotically free; its ultraviolet behaviour is therefore well-defined; and, consequently, one may rigorously define a chiral limit in which the action possesses no current-quark mass terms.  This shows that the right-hand-side (rhs) of Eq.\,\eqref{oGMOR} should involve an expectation-value, evaluated between pion states, of that part of the QCD action which involves the current-quark masses.  It also invites one to consider the action of massless QCD, which, classically, defines a conformally invariant theory.  Such a theory cannot possess a mass-scale; and hence it cannot describe our Universe, whose visible properties are greatly influenced by the existence of a proton whose mass $m_p \approx 1$\,GeV.  A solution of QCD must therefore provide a means by which we can probe and understand the origin of hadron masses, both ground- and excited-states.  Given empirical evidence that $u$- and $d$-quark current-masses are just a few percent of $m_p$, it is plain that explicit chiral symmetry breaking can only be a very small part of the story.

\begin{figure}[t]
\leftline{%
\includegraphics[clip,width=0.30\textwidth]{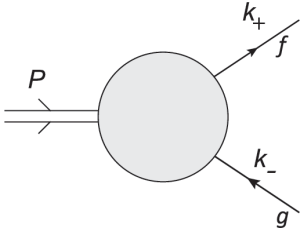}}
\vspace*{-21ex}

\rightline{\parbox{28em}{\caption{\label{figAVvertex}
Momentum flow in a colour-singlet vertex which connects a quark of flavour $f$ and momentum $k_+$ with an antiquark of flavour $\bar g$ and momentum $-k_-$, to form a system with total momentum $P=k_+ - k_-$.  The relative momentum, $k$, is implicitly defined via: $k_+ =  k + \eta P$, $k_- = k - (1-\eta) P$, with $\eta\in[0,1]$; and in a Poincar\'e covariant approach no observable can depend on $\eta$, \emph{i.e}.\ the definition of the relative momentum.
}}}
\end{figure}

It is now sensible to write the explicit form of the AVWGTI:
\begin{eqnarray}
\nonumber \lefteqn{P_\mu \Gamma_{5\mu}^{fg}(k;P) }\\
&&= S_f^{-1}(k_+) i \gamma_5 +  i \gamma_5 S_g^{-1}(k_-) - \, i\,[m_f^{\zeta}+m_g^{\zeta}] \,\Gamma_5^{fg}(k;P)\,,
\label{avwtimN}
\end{eqnarray}
where: $\Gamma_{5\mu}^{fg}(k;P)$ is an inhomogeneous axial-vector vertex, of the type illustrated in Fig.\,\ref{figAVvertex}; $\Gamma_5^{fg}(k;P)$ is an analogous pseudoscalar vertex; $S_{f,g}$ are dressed-quark propagators; and $m_{f,g}^{\zeta}$ are the current-quark masses associated with the two quark flavours, defined at a renormalisation scale $\zeta$.  Actually, each individual term in Eq.\,\eqref{avwtimN} depends on $\zeta$; and yet this identity yields physical results, which cannot.
(\emph{N.B}.\ Ref.\,\cite{Bhagwat:2007ha} discusses the important differences encountered in describing and treating the AVWGTI appropriate to flavourless pseudoscalar mesons.)

The simplest element of Eq.\,\eqref{avwtimN} is the dressed-quark propagator, which can be expressed in a number of equivalent ways:
\begin{subequations}
\label{SgeneralN}
\begin{eqnarray}
 S_f(p) &=&
 -i \gamma\cdot p \,\sigma_V^f(p^2,\zeta^2) + \sigma_S^f(p^2,\zeta^2)\,, \\
\label{SABform}
& = & 1/[i \gamma\cdot p \, A_f(p^2,\zeta^2) + B_f(p^2,\zeta^2)]\,,\\
& = & Z_f(p^2,\zeta^2)/[i\gamma\cdot p + M_f(p^2)]\,.
\end{eqnarray}
\end{subequations}
Importantly, amongst the equivalent functions introduced here, only the mass function, $M_f(p^2) = B_f(p^2,\zeta^2)/A_f(p^2,\zeta^2)$, is independent of the renormalisation point.  The propagator can be obtained from QCD's gap equation, which is a Dyson-Schwinger equation (DSE) \cite{Roberts:1994dr, Cloet:2013jya} that describes how quark propagation is influenced by interactions, \emph{viz}.\ for a quark of flavour $f$,
\begin{subequations}
\label{gendseN}
\begin{eqnarray}
 S_f^{-1}(p) &=& Z_2 \,(i\gamma\cdot p + m_f^{\rm bm}) + \Sigma(p) \,,\\
\Sigma(p) &= & Z_1 \int^\Lambda_{dq}\!\! g^2 D_{\mu\nu}(p-q)\frac{\lambda^a}{2}\gamma_\mu S_f(q) \frac{\lambda^a}{2}\Gamma^f_\nu(q,p) , \rule{1em}{0ex}
\end{eqnarray}
\end{subequations}
where: $D_{\mu\nu}$ is the gluon propagator; $\Gamma^f_\nu$, the quark-gluon vertex; $\int^\Lambda_{dq}$, a symbol that represents a Poincar\'e invariant regularization of the four-dimensional Euclidean integral, with $\Lambda$ the regularization mass-scale; $m_f^{\rm bm}(\Lambda)$, the Lagrangian current-quark bare mass; and $Z_{1,2}(\zeta^2,\Lambda^2)$, respectively, the vertex and quark wave-function renormalisation constants.  Although we have suppressed the renormalisation scale, $\zeta$, in Eqs.\,\eqref{gendseN}, definition of the gap equation is not complete until a renormalisation condition is specified.  A mass-independent scheme is a useful choice and can be implemented by fixing all renormalisation constants in the chiral limit.  (See, \emph{e.g}.\ Ref.\,\cite{Chang:2008ec} and references therein; or Ref.\,\protect\cite{tarrach} for a detailed discussion of renormalisation.)

It is worth noting that the renormalised current-quark mass,
\begin{equation}
\label{mzeta}
m_f^\zeta = Z_m^{-1}(\zeta,\Lambda) \, m^{\rm bm}(\Lambda) = Z_4^{-1} Z_2\, m_f^{\rm bm},
\end{equation}
wherein $Z_4$ is the renormalisation constant associated with the mass term in QCD's Lagrangian, is simply the dressed-quark mass function evaluated at one particular deep-spacelike point, \emph{viz}.\
\begin{equation}
m_f^\zeta = M_f(\zeta^2)\,.
\end{equation}
The renormalisation-group-invariant (RGI) current-quark mass may be inferred via
\begin{equation}
\label{mfhat}
\hat m_f = \lim_{\zeta^2\to\infty} \left[\frac{1}{2}\ln \frac{\zeta^2}{\Lambda^2_{\rm QCD}}\right]^{\gamma_m} M_f(\zeta^2)\,,
\end{equation}
where $\gamma_m = 4/\beta_0$; and the chiral limit is rigorously defined by $\hat m_f = 0$.  Moreover,
\begin{equation}
\label{perturbativemassscaling}
\forall \zeta \gg \Lambda_{\rm QCD}, \;
\frac{m_{f}^\zeta}{m^\zeta_{g}}=\frac{\hat m_{f}}{\hat m_{g}}\,.
\end{equation}

\begin{figure}[t]
\leftline{%
\includegraphics[clip,width=0.45\textwidth]{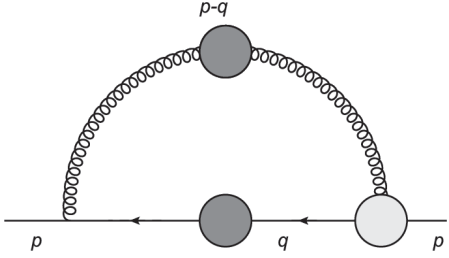}}
\vspace*{-23ex}

\rightline{\parbox{24em}{\caption{\label{FigSigma} Dressed-quark self energy in Eq.\,\protect\eqref{gendseN}.  The kernel is composed from the dressed-gluon propagator (looped-line with dark circle) and the dressed-quark-gluon vertex (light-circle), and the equation is nonlinear owing to the appearance of the dressed-quark propagator (line with dark circle).  This image encodes every imaginable, valid contribution.  (Momentum flows from right-to-left.)}}}

\vspace*{2ex}

\end{figure}

The remaining elements in Eq.\,\eqref{avwtimN} are the amputated axial-vector and pseudoscalar vertices, $\Gamma_{5\mu,5}^{fg}(k;P)$, respectively.  They may both be obtained from an inhomogeneous Bethe-Salpeter equation (BSE) \cite{Salpeter:1951sz}, which is exemplified here using a textbook expression:
\begin{eqnarray}
\nonumber
\lefteqn{[\Gamma_{5\mu}(k;P)]_{tu}}  \\
&&
 = Z_2 [\gamma_5 \gamma_\mu]_{tu} + \int_{dq}^\Lambda [ S(q_+) \Gamma_{5\mu}(q;P) S(q_-) ]_{sr} K_{tu}^{rs}(q,k;P),
\label{bsetextbook}
\end{eqnarray}
in which $K$ is the fully-amputated quark-antiquark scattering kernel, and the indices $r,s,t,u$ denote the colour-, Dirac- and flavour-matrix structure of the elements in the equation.  \emph{N.B}.\ By definition, $K$ does not contain quark-antiquark to single gauge-boson annihilation diagrams, nor diagrams that become disconnected by cutting one quark and one antiquark line: it is two-particle-irreducible.  In quantum field theory, any meson bound-state, constituted from valence $f$ and $\bar g$ quarks and with discrete quantum numbers matching that of the vertex, must appear as a pole in this vertex.  The residue of that pole is proportional to the meson's bound-state Bethe-Salpeter amplitude: there are rare, special circumstances under which the constant of proportionality vanishes \cite{Volkov:1996br, Volkov:1999yi, Holl:2004fr, Holl:2005vu, McNeile:2006qy, Lucha:2006rq, Ballon-Bayona:2014oma, Jiang:2015paa}.

The power of Eq.\,\eqref{avwtimN} now begins to become apparent.  Evidently, no simple connection exists between the kernels of Eqs.\,\eqref{gendseN} and \eqref{bsetextbook}, yet there is an identity which ties their solutions together.  This has numerous, far-reaching consequences; one of which it is important to detail here \cite{Maris:1997hd, Qin:2014vya}.  To proceed, we specialise to the case of light $u$- and $d$-quarks, capitalise on the fact that $P_\mu \Gamma_{5\mu}^{ud}(k;P)$ and $\Gamma_{5}^{ud}(k;P)$ are both $J^{PC}=0^{-+}$ vertices so they must possess the same bound-state poles, and therefore write
\begin{eqnarray}
\nonumber
\Gamma_{5\mu}^{ud}(k;P) &\stackrel{P^2+m_{\pi}^2\simeq 0}{=} &
\gamma_5 \left[ \gamma_\mu \, F_A(k;P) + k_\mu \gamma\cdot k \, G_A(k;P) \right] \\
\nonumber
&& \left. - \sigma_{\mu\nu} k_\nu \, H_A(k;P)\right] + \tilde \Gamma_{5\mu}^{ud}(k;P) \\
\label{AVvtx}
&& + \frac{2 f_{\pi} P_\mu}{P^2+m_{\pi}^2}\Gamma_\pi(k;P) \,,\\
\nonumber
\Gamma_{5}^{ud}(k;P) &\stackrel{P^2+m_{\pi}^2\simeq 0}{=} &
i\gamma_5 \left[E_5(k;P) + \gamma\cdot P \, F_5(k;P) + \gamma\cdot k k\cdot P \, G_5(k;P) \right.\\
&& \left. + \sigma_{\mu\nu}k_\mu P_\nu H_5(k;P) \right]
\label{5vtx}
+ \frac{2 \rho_{\pi} P_\mu}{P^2+m_{\pi}^2}\Gamma_\pi(k;P)\,,
\end{eqnarray}
where $E_5$, $F_{A,5}$, $G_{A,5}$, $H_{A,5}$ are regular in the neighbourhood $P^2+m_{\pi}^2\simeq 0$, $\tilde P_\mu \Gamma_{5\mu,5}^{ud}\sim {\rm O}(P^2+m_{\pi}^2)$,  and the putative bound-state's Bethe-Salpeter amplitude is \cite{LlewellynSmith:1969az}:
\begin{eqnarray}
\nonumber \Gamma_\pi(k;P) = i \gamma_5 \left[ E_\pi(k;P)  \right.\\
\left. + \gamma\cdot P \, F_\pi(k;P) + \gamma\cdot k k\cdot P \, G_\pi(k;P) + \sigma_{\mu\nu} k_\mu P_\nu\, H_\pi(k;P) \right]\,. \label{Gammapi}
\end{eqnarray}
For use hereafter, we note that it is often useful to characterise the scalar functions in Eq.\,\eqref{Gammapi} by their Chebyshev moments:
\begin{equation}
\label{Chebyshev}
{\mathpzc T}^{(m)}(k^2;P^2) = \frac{2}{\pi} \int_{-1}^{1} \! dx\, \sqrt{1-x^2} \, U_m(x) {\mathpzc T}(k;P) \,,
\end{equation}
where $x=k\cdot P/\sqrt{k^2 P^2}$ and $\{U_m(x),m=0,1,\ldots \}$ is the set of Chebyshev polynomials of the second kind.

Consider now the chiral limit: $\hat m_u = 0 = \hat m_d$, so that final term in Eq.\,\eqref{avwtimN} vanishes, and suppose that $m_\pi^2=0$ in this limit, then the AVWGTI entails \cite{Maris:1997hd, Qin:2014vya} the following array of Goldberger-Treiman relations, involving the chiral-limit solutions for the functions in Eq.\,\eqref{SABform}:
\begin{subequations}
\label{gGTrelations}
\begin{eqnarray}
f_\pi^0 E_{\pi}^0(y,w=0;P^2=0) &=& B_0(y) \,, \label{BGTrelation}\\
F_A^0(y,w=0;P^2=0) + 2 f_\pi^0 F_\pi^0(y,w=0;P^2=0) &=& A_0(y) \,, \\
G_A^0(y,w=0;P^2=0) + 2 f_\pi^0 G_\pi^0(y,w=0;P^2=0) &=& 2 A_0^\prime(y) \,, \\
H_A^0(y,w=0;P^2=0) + 2 f_\pi^0 H_\pi^0(y,w=0;P^2=0) &=& 0 \,,
\end{eqnarray}
\end{subequations}
where $y=k^2$, $w=k\cdot P$.  In perturbation theory, $B(k^2) \equiv 0$ in the chiral limit.  The appearance of a $B(k^2)$-nonzero solution of (\ref{gendseN}) in the chiral limit signals DCSB: one has {\it dynamically generated} a running quark mass in the absence of a seed-mass.  Eqs.\,(\ref{gGTrelations}) show that when chiral symmetry is dynamically broken: 1) the homogeneous, isovector, pseudoscalar BSE has a massless, $P^2=0$, solution; 2) the Bethe-Salpeter amplitude for the massless bound-state has a term proportional to $\gamma_5$ alone, with $E_\pi(k;0)$ completely determined by the scalar part of the quark self-energy, in addition to other pseudoscalar Dirac structures, $F_\pi$, $G_\pi$ and $H_\pi$, which are nonzero in general and play a crucial role in the large-$Q^2$ behaviour of pseudoscalar meson elastic form factors \cite{Maris:1998hc, GutierrezGuerrero:2010md, Chen:2012txaS}; and 3) the axial-vector vertex is dominated by the associated pole on $P^2\simeq 0$.  Notably, this last statement is an expression of PCAC.  Moreover, the converse is also true; namely, 1)-3) entail DCSB.

It is thus seen that \\[-2ex]
\hspace*{4em}\parbox[t]{0.8\textwidth}{\flushleft \emph{in the chiral limit, DCSB is a sufficient and necessary condition for the appearance of a massless pseudoscalar bound-state that dominates the axial-vector vertex on $P^2\simeq 0$ and whose constituents are described by a momentum-dependent mass-function, which may be arbitrarily large}.}
\medskip

Proceeding to the case $\hat m_u \neq 0 \neq \hat m_d$, Eqs.\,\eqref{avwtimN}, \eqref{AVvtx}, \eqref{5vtx} ensure that in the neighbourhood of any $0^{-+}$ pole located at $P^2+m_\pi^2=0$:
\begin{equation}
\label{gGMOR}
f_\pi m_\pi^2 = (m_u^\zeta +m_d^\zeta) \rho_\pi^\zeta\,,
\end{equation}
along with generalised versions of Eqs.\,\eqref{gGTrelations} \cite{Qin:2014vya}.  It will immediately be apparent that with Eq.\,\eqref{gGMOR} one has recovered Eq.\,\eqref{oGMOR}.  Thus DCSB, expressed via the appearance of an arbitrarily large dressed-quark mass-function, is necessary and sufficient to ensure the peculiar behaviour of pseudo-Nambu-Goldstone-boson masses.

There are now a few obvious questions to ask: (1) what does it take to generate a nonzero running mass-function in the chiral limit; and (2) what are the quantities $f_\pi$, $\rho_\pi^\zeta$?  The first of these is equivalent to asking how does the self-energy depicted in Fig.\,\ref{FigSigma} generate \emph{mass from nothing}, something which is impossible in the classical theory.  The answer lies in the fact that Fig.\,\ref{FigSigma} is a deceptively simply picture.  It actually corresponds to a countable infinity of diagrams, all of which can potentially contribute, and may also express nonperturbative contributions that might not have a diagrammatic representation at all.  To provide a context, quantum electrodynamics, an Abelian gauge theory, has 12\,672 diagrams at order $\alpha^5$ in the computation of the electron's anomalous magnetic moment \cite{Aoyama:2012wj}.  Owing to its foundation in the non-Abelian group $SU(3)$, the analogous perturbative computation of a quark's anomalous chromomagnetic moment has many more diagrams at this order in the strong coupling.  The number of diagrams represented by the self energy in Fig.\,\ref{FigSigma} grows equally rapidly, \emph{i}.\emph{e}.\ combinatorially with the number of propagators and vertices used at a given order.  Indeed, proceeding systematically, a computer will very quickly generate the first diagram in which the number of loops is so great that it is simply impossible to calculate in perturbation theory: impossible in the sense that we don't yet have the mathematical capacity to solve the problem.

\begin{figure}[t]
\centerline{\includegraphics[clip,width=0.5\textwidth]{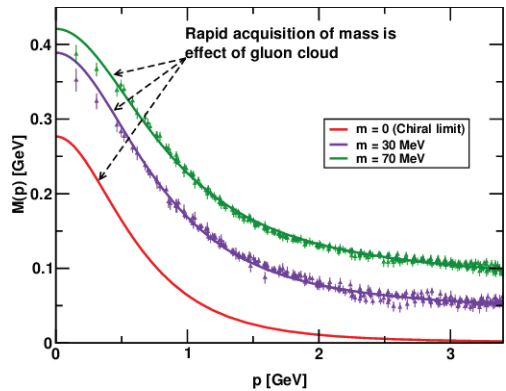}}
\caption{\label{gluoncloud}
Dressed-quark mass function, $M(p)$ in Eq.\,(\protect\ref{SgeneralN}): \emph{solid curves} -- DSE results, explained in Refs.\,\protect\cite{Bhagwat:2003vw,Bhagwat:2006tu}, ``data'' -- numerical simulations of lattice-regularised QCD (lQCD) \protect\cite{Bowman:2005vx}.  (NB.\ $m=70\,$MeV is the uppermost curve and current-quark mass decreases from top to bottom.)  The current-quark of perturbative QCD evolves into a constituent-quark as its momentum becomes smaller.  The constituent-quark mass arises from a cloud of low-momentum gluons attaching themselves to the current-quark.  This is DCSB: an essentially nonperturbative effect that generates a quark \emph{mass} \emph{from nothing}; namely, it occurs even in the chiral limit.}
\end{figure}

Each of the diagrams which contributes to $M(p^2)$ in a weak-coupling expansion of Fig.\,\ref{FigSigma} is multiplied by the current-quark mass, $\hat m$.  Plainly, any finite sum of diagrams must therefore vanish as $\hat m\to 0$.  However, with \emph{infinitely many} diagrams the situation might be very different: one has ``$ 0 \times \infty$,'' a product whose limiting value is contingent upon the cumulative magnitude of each term in the sum.  Consider therefore the behaviour of $M(p^2)$ at large $p^2$.  QCD is asymptotically free \cite{Politzer:2005kc, Gross:2005kv, Wilczek:2005az}.  Hence, on this domain, each of the regularised loop diagrams must individually evaluate to a small number whose value depends on just how large is the coupling.  It will not be surprising, therefore, to learn that for a monotonically-decreasing running-coupling, $\alpha_S(k^2)$, there is a critical value of $\alpha_S(0)$ above which the magnitude of the sum of infinitely many diagrams is sufficient to balance the linear decrease of $\hat m\to 0$, so that the answer is nonzero and finite in this limit, \emph{viz}.
\begin{equation}
\exists\, \alpha_S^c(0) \; |\; \forall \alpha_S(0) >\alpha_S^c(0),
M_0(p^2):= \lim_{\hat m\to 0} M(p^2;\hat m) \neq 0\,.
\end{equation}
The internal consistency of QCD appears to guarantee that the limit is always finite.  In fact, QCD generates a mass-function of the type depicted in Fig.\,\ref{gluoncloud}.

The scale of the mass-function in Fig.\,\ref{gluoncloud} is striking: even the nontrivial chiral limit solution, which cannot exist perturbatively, reaches a value of roughly 300\,MeV at infrared momenta, \emph{i.e}.\ one dressed-quark possesses one-third of the proton's mass.  It follows that the relationship in Eq.\,\eqref{perturbativemassscaling} is broken by nonperturbative dynamics, so that
\begin{equation}
\frac{m_{f}^{\zeta=p^2}}{m^{\zeta=p^2}_{g}}=\frac{M_{f}(p^2)}{M_{g}(p^2)}
\end{equation}
is not independent of $p^2$: in the infrared, \emph{i.e}.\ $\forall p^2 \lesssim 2\,{\rm GeV}^2 =: \Lambda_\chi^2$, it expresses a ratio of constituent-like quark masses, which, for light quarks, are two orders-of-magnitude larger than their current-masses and nonlinearly related to them \cite{Flambaum:2005kc, Holl:2005st}.  Plainly, DCSB is the primary source of the proton's mass and hence of the vast bulk of visible matter in the Universe.

The appearance of a dynamically-generated nonzero mass-function in the solution of QCD's chiral-limit one-quark problem has additional fascinating consequences, \emph{e.g}.\ Eqs.\,\eqref{gGTrelations} in general, and Eq.\,\eqref{BGTrelation} in particular, acquire an extraordinary character. These equations mean that the pseudoscalar two-body problem is solved, well-nigh completely and without additional effort, once the solution to the one-body dressed-quark problem is known; and, furthermore, that the quark-level Goldberger-Treiman relation in Eq.\,\eqref{BGTrelation} is the most basic expression of Goldstone's theorem in QCD, \emph{viz}.\\[-2ex]
\hspace*{4em}\parbox[t]{0.8\textwidth}{\flushleft \emph{Goldstone's theorem is fundamentally an expression of equivalence between the one-body problem and the two-body problem in QCD's colour-singlet pseudoscalar channel}.}
\medskip

\hspace*{-\parindent}Consequently, pion properties are an almost direct measure of the mass function depicted in Fig.\,\ref{gluoncloud}; and the reason a pion is massless in the chiral limit is simultaneously the explanation for a proton mass of around 1\,GeV.  Thus, enigmatically, properties of the nearly-massless pion are the cleanest expression of the mechanism that is responsible for almost all the visible mass in the Universe.

We are left to answer the second question posed after Eq.\,\eqref{gGMOR}; namely, what are the two nonperturbatively generated constants in the GMOR relation or what are the residues of the pion pole in the axial-vector and pseudsocalar vertices?  A little additional analysis of the quark-antiquark scattering matrix reveals \cite{Maris:1997hd}:
\begin{subequations}
\label{fpigen}
\begin{eqnarray}
i f_{\pi} P_\mu  &=& \langle 0 | \bar d \gamma_5 \gamma_\mu u |\pi\rangle \\
&=& Z_2\; {\rm tr}_{\rm CD}
\int_{dq}^\Lambda i\gamma_5\gamma_\mu S_{u}(q_+) \Gamma_{\pi}(q;P) S_{d}(q_-)\,, \end{eqnarray}
\end{subequations}
\vspace*{-5ex}

\begin{subequations}
\label{rhogen}
\begin{eqnarray}
i\rho_{\pi}^\zeta &=& -\langle 0 | \bar d i\gamma_5 u |\pi \rangle \\
&=& Z_4\; {\rm tr}_{\rm CD}
\int_{dq}^\Lambda \gamma_5 S_{u}(q_+) \Gamma_{\pi}(q;P) S_{d}(q_{-}) \,,
\end{eqnarray}
\end{subequations}
where the trace is over colour and spinor indices.  The expressions in Eqs.\,\eqref{fpigen} might be recognised as those defining the pion's leptonic decay constant, which, owing to the factor $Z_2$, is a pseudovector projection of the pion's Bethe-Salpeter wave function onto the origin in configuration space that is both gauge-invariant and independent of the renormalisation scale, $\zeta$.   Equations~\eqref{rhogen} express an analogous ``decay constant'', \emph{viz}.\ a pseudoscalar projection of the pion's wave function.  Owing to the presence of $Z_4$, this quantity is also gauge invariant and evolves with renormalisation scale in precisely the manner required to ensure the rhs of Eq.\,\eqref{gGMOR} is $\zeta$-independent.

Notwithstanding these features, Eq.\,\eqref{gGMOR} is not a form of the GMOR relation this is commonly seen in textbooks. The connection may be understood by introducing
\begin{equation}
\label{kappazeta}
\kappa^\zeta_{\pi} := f_{\pi} \rho_{\pi}^\zeta =: [\chi_{\pi}^\zeta]^3\,,
\end{equation}
for then Eq.\,\eqref{gGMOR} can be written
\begin{equation}
\label{gGMOR2}
f_{\pi}^2 m_{\pi}^2 = [m_{u}^\zeta +m_{d}^\zeta]\, \kappa^\zeta_{\pi}
=: [\hat m_{u} + \hat m_{d}]\, \hat \kappa_{\pi},
\end{equation}
in which form $\kappa^\zeta_{\pi}$ is seen to play the role of what is commonly called the chiral condensate.  Notably, Eqs.\,\eqref{gGMOR} and \eqref{gGMOR2} are special cases of an exact mass formula in QCD that is valid for arbitrarily small or large current-quark masses and for both ground- and excited-state pseudoscalar mesons \cite{Holl:2004fr}; and, owing to this feature, Eq.\,\eqref{kappazeta} has come to be recognised as an in-hadron condensate \cite{Maris:1997tm}.  That connection is cemented by reiterating an identity, proved elsewhere \cite{Maris:1997hd}:
\begin{equation}
\lim_{\hat m\to 0} \kappa^\zeta_{\pi}
= - \lim_{\hat m\to 0} f_\pi \langle 0 | \bar q i\gamma_5 q |\pi \rangle
=
Z_4 \, {\rm tr}_{\rm CD}\int_{dq}^\Lambda S^0(q;\zeta) =  -\langle \bar q q \rangle_\zeta^0\,;
\label{qbqpiqbq0}
\end{equation}
namely, the so-called vacuum quark condensate is, in fact, the chiral-limit value of the in-meson condensate, \emph{viz}.\ it describes a property of the chiral-limit pseudoscalar meson.  This condensate is therefore no more a property of the ``vacuum'' than the pseudoscalar meson's chiral-limit leptonic decay constant.  Moreover, given that Eq.\,\eqref{qbqpiqbq0} is an identity in QCD, any veracious calculation of $\langle \bar q q \rangle_\zeta^0$ is the computation of a gauge-invariant property of the pion's wave-function.
Extensive discussion of the consequences deriving from this change in perspective is presented in Refs.\,\cite{Brodsky:2009zd, Brodsky:2010xf, Chang:2011mu, Brodsky:2012ku, Cloet:2013jya}.

It is worth closing this section with some additional remarks on confinement, given the role it plays in forcing a changed view on the nature of condensates and, indeed, the integral part that confinement is likely to play in determining whether or not QCD is a mathematically well-defined relativistic quantum field theory in four dimensions.  One aspect of the Yang-Mills millennium problem \cite{Jaffe:Clay} is to prove that pure-gauge QCD possesses a mass-gap $\Delta>0$.  There is strong evidence supporting this conjecture, found especially in the fact that numerical simulations of lattice-regularised QCD (lQCD) predict $\Delta \gtrsim 1.5\,$GeV \cite{McNeile:2008sr}.  This sharpens the conundrum we presented in the Introduction: with $\Delta^2/m_\pi^2 \gtrsim 100$, can the mass-gap in pure Yang-Mills really play any role in understanding confinement when DCSB, driven by the same dynamics, ensures the existence of an almost-massless strongly-interacting excitation in our Universe?  If the answer is not \emph{no}, then it must at least be that one cannot claim to provide a pertinent understanding of confinement without simultaneously explaining its connection with DCSB.  The pion must play a critical role in any explanation of confinement in the Standard Model; and any discussion that omits reference to the pion's role is \emph{practically} \emph{irrelevant}.

This perspective is canvassed elsewhere \cite{Cloet:2013jya} and can be used to argue that the potential between infinitely-heavy quarks measured in numerical simulations of quenched lQCD -- the so-called static potential \cite{Wilson:1974sk} -- is disconnected from the question of confinement in our Universe.  This is because light-particle creation and annihilation effects are essentially nonperturbative in QCD, so it is impossible in principle to compute a quantum mechanical potential between two light quarks \cite{Bali:2005fuS, Prkacin:2005dc, Chang:2009ae}.  It follows that there is no flux tube in a Universe with light quarks and consequently that the flux tube is not the correct paradigm for confinement.

As we have highlighted, DCSB is the key here.  It ensures the existence of pseudo-Nambu-Goldstone modes; and in the presence of these modes, no flux tube between a static colour source and sink can have a measurable existence.  To verify this statement, consider such a tube being stretched between a source and sink.  The potential energy accumulated within the tube may increase only until it reaches that required to produce a particle-antiparticle pair of the theory's pseudo-Nambu-Goldstone modes.  Simulations of lQCD show \cite{Bali:2005fuS, Prkacin:2005dc} that the flux tube then disappears instantaneously along its entire length, leaving two isolated colour-singlet systems.  The length-scale associated with this effect in QCD is $r_{\not\sigma} \simeq (1/3)\,$fm and hence if any such string forms, it would dissolve well within a hadron's interior.

An alternative realisation associates confinement with dramatic, dynamically-driven changes in the analytic structure of QCD's propagators and vertices.  That leads coloured $n$-point functions to violate the axiom of reflection positivity and hence forces elimination of the associated excitations from the Hilbert space associated with asymptotic states \cite{GJ81}.  This is certainly a sufficient condition for confinement \cite{Stingl:1985hx, Krein:1990sf, Hawes:1993ef, Roberts:1994dr}.  It should be noted, however, that the appearance of such alterations when analysing some truncation of a given theory does not mean that the theory itself is truly confining: unusual spectral properties can be introduced by approximations, leading to a truncated version of a theory which is confining even though the complete theory is not, \emph{e.g}.\ Refs.\,\cite{Krein:1993jb, Bracco:1993cy}.  Notwithstanding exceptions like these, a computed violation of reflection positivity by coloured functions in a veracious treatment of QCD does express confinement.  Moreover, via this mechanism, it is achieved as the result of an essentially dynamical process.  Figure~\ref{gluoncloud} highlights that quarks acquire a running mass distribution in QCD; and, as will become clear, this is also true of gluons (see, \emph{e}.\emph{g}.\ Refs.\,\cite{Aguilar:2008xm, Aguilar:2009nf, Boucaud:2011ugS, Pennington:2011xs, Ayala:2012pb, Binosi:2014aea}).  The generation of these masses leads to the emergence of a length-scale $\varsigma \approx 0.5\,$fm, whose existence and magnitude is evident in all existing studies of dressed-gluon and -quark propagators, and which characterises the dramatic change in their analytic structure that we have just described.  In models based on such features \cite{Stingl:1994nk}, once a gluon or quark is produced, it begins to propagate in spacetime; but after each ``step'' of length $\varsigma$, on average, an interaction occurs so that the parton loses its identity, sharing it with others.  Finally a cloud of partons is produced, which coalesces into colour-singlet final states.  This picture of parton propagation, hadronisation and confinement can be tested in experiments at modern and planned facilities \cite{Accardi:2009qv, Dudek:2012vr, Accardi:2012qutSd}.

\section{Continuum Bound-State Problem}
\label{secCBSP}
As indicated in the Introduction, we will chiefly herein review predictions for pion elastic and transition form factors that were obtained using a continuum approach to the two valence-body bound-state problem in relativistic quantum field theory.  In order to enable an appraisal of those predictions, it is appropriate to briefly review the status of such studies.

A natural framework within which to express and preserve the symmetries relevant to the pion is provided by QCD's DSEs \cite{Roberts:2015lja}.  In fact, all bound-state problems may be formulated this way.  The DSEs are a tower of coupled integral equations, examples of which are Eqs.\,\eqref{gendseN}, \eqref{bsetextbook}.  They are well-suited to the study of problems in QCD because their simplest use is as a generating tool for perturbation theory.  Owing to asymptotic freedom, that materially reduces model dependence in sound applications because the interaction kernel in each DSE is known for all momenta within the perturbative domain, \emph{i.e}.\ $k^2\gtrsim 2\,$GeV$^2$: any model need then only be defined with reference to the long-range behaviour of the kernels.  This is good because DSE solutions are Schwinger functions, \emph{i.e}.\ propagators and vertices; and since all cross-sections are built from Schwinger functions, the approach connects observables with the long-range behaviour of the theory's running coupling and masses.  Consequently, feedback between theoretical predictions and experimental tests can then refine the statements and lead to an understanding of these fundamental quantities.  Those predictions are wide-ranging because the DSEs provide a nonperturbative, continuum approach to hadron physics and can therefore address questions pertaining to, \emph{e.g}.: the gluon- and quark-structure of hadrons; and the roles of emergent phenomena -- confinement and DCSB -- and the connections between them.

As a collection of coupled equations, a tractable problem is only obtained once a DSE truncation scheme is specified.  It is unsurprising that the best known procedure is the weak coupling expansion, which reproduces every diagram in perturbation theory.  That scheme is systematic and valuable in the analysis of large momentum transfer phenomena; but it precludes any possibility of obtaining nonperturbative information.  A systematic, symmetry-preserving scheme applicable to hadron bound-states is described in Refs.\,\cite{Munczek:1994zz, Bender:1996bb, Binosi:2016rxzd}.  The procedure generates a Bethe-Salpeter equation from the kernel of any gap equation whose diagrammatic content is known and thereby guarantees, \emph{inter} \emph{alia}, that Eq.\,\eqref{avwtimN} and kindred identities can always be preserved in any treatment of composite operators in QCD.  The mere existence of this scheme enabled the proof of exact nonperturbative results (see, \emph{e.g}.\ Refs.\,\cite{Bashir:2012fs, Chang:2012cc}) and it remains the most widely used today.

The leading-order term in the procedure of Refs.\,\cite{Munczek:1994zz, Bender:1996bb, Binosi:2016rxzd} is the rainbow-ladder (RL) truncation.  It is widely used and known to be accurate for light-quark ground-state vector- and isospin-nonzero-pseudoscalar-mesons \cite{Maris:2003vk, Chang:2011vu, Bashir:2012fs, Cloet:2013jya}, and properties of ground-state octet and decuplet baryons \cite{Eichmann:2011ej, Chen:2012qr, Segovia:2013rca, Segovia:2013uga}, because corrections in these channels largely cancel owing to the parameter-free preservation of relevant WGT identities ensured by this scheme.  However, higher-order contributions do not typically cancel in other channels \cite{Roberts:1996jxS, Bender:2002as, Bhagwat:2004hn}.  Hence studies based on the RL truncation, or low-order improvements thereof, usually provide poor results for light-quark scalar- and axial-vector-mesons \cite{Burden:1996nh, Watson:2004kd, Maris:2006ea, Cloet:2007piS, Fischer:2009jm, Krassnigg:2009zh},
exhibit gross sensitivity to model parameters for tensor-mesons \cite{Krassnigg:2010mh} and excited states \cite{Holl:2004fr, Holl:2004un, Qin:2011dd, Qin:2011xq}, and are unrealistic for heavy-light systems \cite{Nguyen:2010yh, Rojas:2014aka, Gomez-Rocha:2015qga}.

These difficulties are surmounted in a recently developed truncation scheme \cite{Chang:2009zb}, which is beginning to have a material impact.  The new scheme enables the use of far more sophisticated kernels for the gap and Bethe-Salpeter equations, which include, \emph{e.g}.\ Dirac vector$\otimes$vector and scalar$\otimes$scalar quark-antiquark interactions, and overcome the weaknesses of RL truncation in all channels studied thus far.  The new technique, too, is symmetry preserving; but it has additional strengths, \emph{e.g}.\ the capacity to express DCSB nonperturbatively in the integral equations connected with bound-states.  Owing to this feature, the new scheme is described as the ``DCSB-improved'' or ``DB'' truncation.  It preserves successes of the RL truncation; but has also enabled elucidation of many novel nonperturbative features of QCD.  For instance, the existence of dressed-quark anomalous chromo- and electro-magnetic moments \cite{Chang:2010hb} and the key role they play in determining observable quantities \cite{Chang:2011tx}; elucidation of the causal connection between DCSB and the splitting between vector and axial-vector mesons \cite{Chang:2011ei}; and the impact of that splitting on the baryon spectrum \cite{Chen:2012qr}.  Furthermore, as will subsequently be highlighted, development of the DB truncation has enabled a crucial step toward the \emph{ab initio} prediction of hadron observables in continuum-QCD.

The connection between the existence of Nambu-Goldstone modes and a dynamically-generated dressed-quark mass was discussed in connection with Figs.\,\ref{FigSigma} and \ref{gluoncloud}.  The propagation of gluons, too, is described by a gap equation \cite{Aguilar:2008xm, Aguilar:2009nf, Boucaud:2011ugS, Pennington:2011xs, Ayala:2012pb, Binosi:2014aea}; and its solution shows that gluons are cannibals: they are a particle species whose members become massive by eating each other!  The associated gluon mass function, $m_g(k^2)$, is monotonically decreasing with increasing $k^2$ and recent work \cite{Binosi:2014aea} has established that
\begin{equation}
\label{gluonmassEq}
m_g(k^2=0) \approx 0.5\,{\rm GeV}.
\end{equation}
The value of the mass-scale in Eq.\,\eqref{gluonmassEq} is \emph{natural} in the sense that it is commensurate with but larger than the value of the dressed light-quark mass function at far infrared momenta: $M(0)\approx 0.3\,$GeV (see Fig.\,\ref{gluoncloud}).  Moreover, the mass term appears in the transverse part of the gluon propagator, hence gauge-invariance is not tampered with; and the mass function falls as $1/k^2$ for $k^2\gg m_g(0)$ (up to logarithmic corrections), so the gluon mass is invisible in perturbative applications of QCD: it has dropped to less-than 5\% of it's infrared value by $k^2=4\,$GeV$^2$.

Crucially, gauge boson cannibalism presents a new physics frontier within the Standard Model.  Asymptotic freedom means that the ultraviolet behaviour of QCD is controllable.  At the other extreme, dynamically generated masses for gluons and quarks entail that QCD creates its own infrared cutoffs.  Together, these effects eliminate both the infrared and ultraviolet problems that typically plague quantum field theories and thereby make reasonable the hope that QCD is nonperturbatively well defined, \emph{viz}.\ that the millennium problem \cite{Jaffe:Clay} does have a solution.  The presence of dynamically-generated gluon and quark mass-scales must have many observable consequences, too, and hence can be checked experimentally.  For example, one may plausibly conjecture that dynamical generation of an infrared gluon mass-scale leads to saturation of the gluon parton distribution function at small Bjorken-$x$ within hadrons.  This could be checked via computations of gluon distribution functions, using such solutions of the gluon gap equation in hadron bound-state equations.  The possible emergence of this phenomenon stirs great interest; and it is a key motivation in plans to construct an EIC that would be capable of producing a precise empirical understanding of collective behaviour amongst gluons \cite{Accardi:2012qutSd}.

As Eqs.\,\eqref{gendseN}, \eqref{bsetextbook} make apparent, one must possess detailed information about the interactions between quarks and gluons at all momentum scales in order to solve the continuum bound-state problem.  There are two common methods for determining this information: the top-down approach, which works toward an \textit{ab initio} computation of the interaction via direct analysis of the gauge-sector gap equations; and the bottom-up scheme, which aims to infer the interaction by fitting data within a well-defined truncation of those equations in the matter sector that are relevant to bound-state properties.  These two approaches have recently been united \cite{Binosi:2014aea} by a demonstration that the RGI running-interaction predicted by contemporary analyses of QCD's gauge sector coincides with that required in order to describe ground-state hadron observables using the DB truncation.

\begin{figure}[t]
\rightline{\begin{minipage}[t]{0.8\textwidth}
\begin{minipage}{0.7\textwidth}
\centerline{\includegraphics[clip,width=0.97\textwidth]{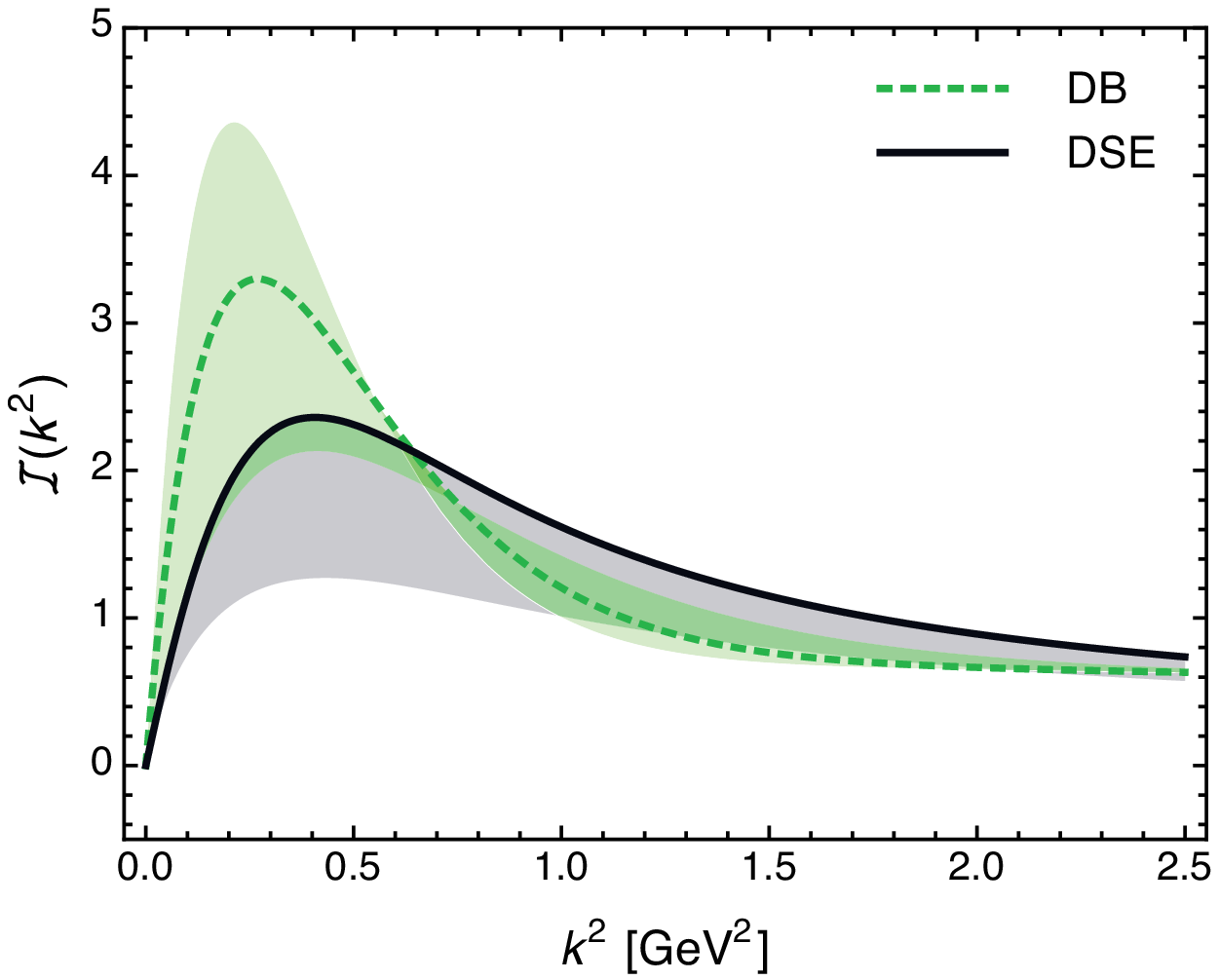}}
\end{minipage}
\begin{minipage}{0.705\textwidth}
\centerline{\includegraphics[clip,width=0.97\textwidth]{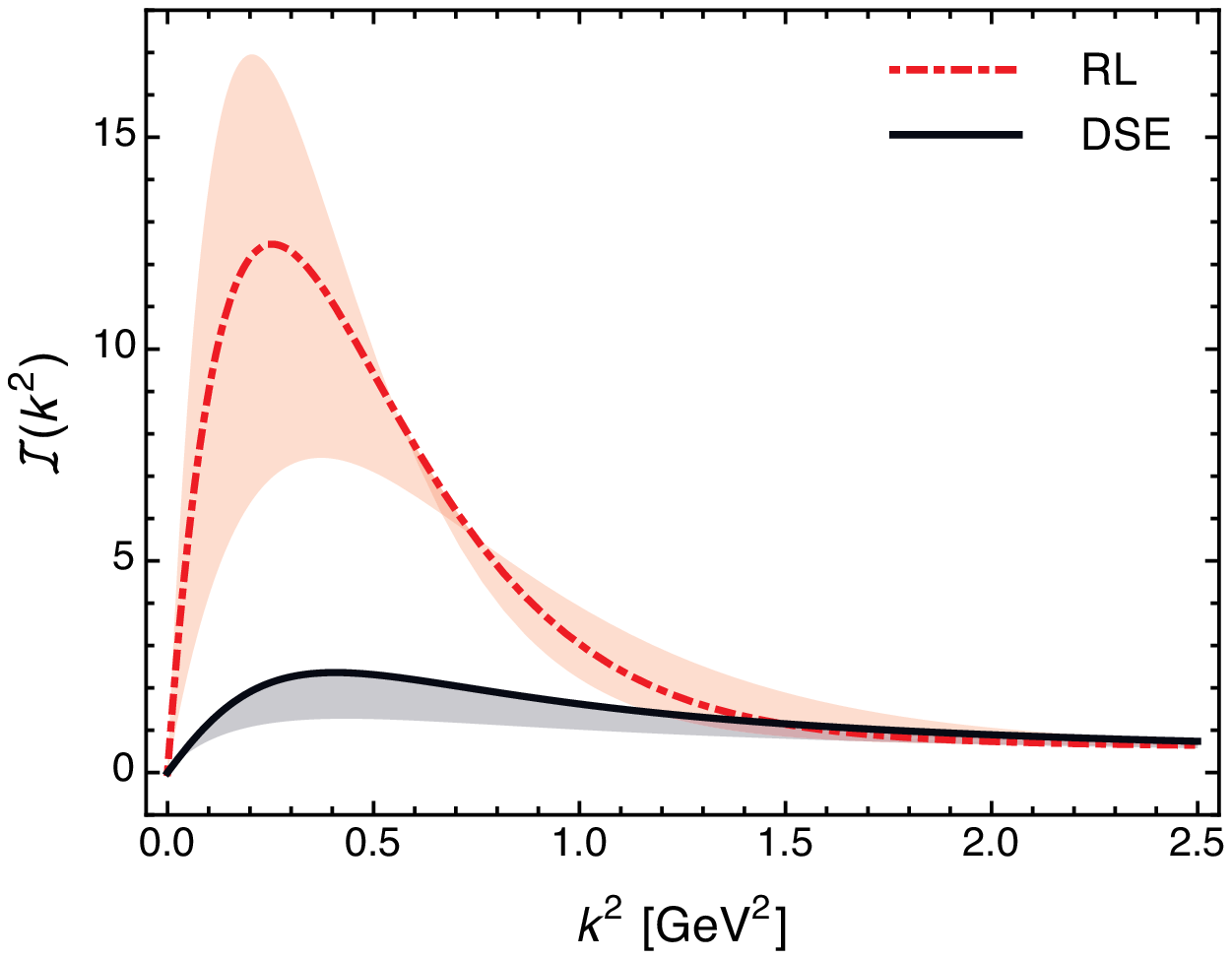}}
\end{minipage}
\end{minipage}}
\caption{Comparison between top-down results for the gauge-sector interaction and  with those obtained using the bottom-up approach based on hadron physics observables.
\textbf{Upper panel} -- \emph{solid curve} within grey band, top-down result for the RGI interaction; and \emph{dashed curve} within \emph{pale-green band}, advanced bottom-up result obtained in the DB truncation.
\textbf{Lower panel} -- \emph{solid curve} within grey band, top-down result for the RGI interaction, as in the left panel; and \emph{dot-dashed curve} within \emph{pale-red band}, bottom-up result obtained in the RL truncation.
In all cases, the bands denote the existing level of theoretical uncertainty in extraction of the interaction.
All curves are identical on the perturbative domain: $k^2>2.5\,$GeV$^2$.
(Figures provided by D.~Binosi, modelled after those in Ref.\,\cite{Binosi:2015xqk}.)
\label{TopBottom}}
\end{figure}

These observations are illustrated by Fig.\,\ref{TopBottom}.  The upper-panel shows that the top-down RGI interaction and the DB-truncation bottom-up interaction agree within existing theoretical error; namely, the interaction predicted by modern analyses of QCD's gauge sector is in good agreement with that required for a veracious description of measurable hadron properties using the most sophisticated matter-sector gap and Bethe-Salpeter kernels available today.  This is a remarkable result, given that there had previously been no serious attempt at communication between practitioners from the top-down and bottom-up hemispheres of continuum-QCD.  It bridges a gap that had lain between nonperturbative continuum-QCD and the \emph{ab initio} prediction of bound-state properties.

The bottom-panel in Fig.\,\ref{TopBottom} shows that the interaction inferred using a modern RL truncation \cite{Qin:2011ddS, Qin:2011xqS} has the correct shape; but it is too large in the infrared.  This is because the RL truncation suppresses all effects associated with DCSB in the kernels of the gap and Bethe-Salpeter equations \emph{except} those expressed in the running coupling itself, and therefore a description of hadronic phenomena can only be achieved by overmagnifying the gauge-sector interaction strength at infrared momenta.  A similar conclusion is drawn elsewhere \cite{Rojas:2014tya}.  It follows that whilst the RL truncation supplies a useful computational link between QCD's gauge sector and measurable hadron properties, the model interaction it delivers should neither be misconstrued nor misrepresented as a pointwise-accurate representation of ghost-gluon dynamics.  Notwithstanding this, the judicious use of RL truncation and the careful interpretation of its results can still be a valuable tool for hadron physics phenomenology.

\section{Leading-Twist PDA}
\label{secLTPDA}
The light-front wave-function of an interacting quantum system, $\varphi(x)$, provides a connection between dynamical properties of the underlying relativistic quantum field theory and notions familiar from nonrelativistic quantum mechanics.  In particular, although particle number conservation is generally lost in relativistic quantum field theory, $\varphi(x)$ has a probability interpretation.  It can therefore translate features that arise purely through the infinitely-many-body nature of relativistic quantum field theory into images whose interpretation seems more straightforward \cite{Keister:1991sb, Brodsky:1997de, Chang:2013pqS}.  This would be very useful if realised in connection with confinement and DCSB.

As described in Sec.\,\ref{secTheoryA}, pion properties are a particularly clear manifestation of DCSB.  This fact is highlighted by the quark-level Goldberger-Treiman relations in Eqs.\,\eqref{gGTrelations}, which entail that the $x$-dependence of $\varphi_\pi(x;\zeta^2)$ must provide a nearly pure expression of this crucial emergent phenomena in the Standard Model; and hence the calculation of $\varphi_\pi(x;\zeta^2)$ provides a means by which to expose DCSB in a wave function with quantum mechanical characteristics.

The pion bound-state problem derived from Eqs.\,\eqref{gendseN}, \eqref{bsetextbook} was solved in Ref.\,\cite{Chang:2013pqS}, using both the RL- and DB-truncations with the matching interaction drawn from Fig.\,\ref{TopBottom}.  Those solutions were then used to construct the Poincar\'e-covariant Bethe-Salpeter wave function for the pion:
\begin{equation}
\chi_\pi(q;P) = S_u(q_+) \Gamma_\pi(q_{+-};P) S_d(q_-)\,,
\label{chipi}
\end{equation}
as determined in the given truncation, where $q_{+-}= [q_++q_-]/2$ and $\Gamma_\pi$ is the Bethe-Salpeter amplitude in Eq.\,\eqref{Gammapi}.  With the wave function in hand, the PDA in Eq.\,\eqref{pionUV} was calculated via its definition as the projection of $\chi_\pi(q;P)$ onto the light-front:
\begin{equation}
f_\pi\, \varphi_\pi(x;\zeta^2) = {\rm tr}_{\rm CD}
Z_2 \! \int_{dq}^\Lambda \!\!
\delta(n\cdot q_+ - x \,n\cdot P) \,\gamma_5\gamma\cdot n\, \chi_\pi(q;P)\,.
\label{pionPDA}
\end{equation}
Here $n$ is a light-like four-vector, $n^2=0$; $P$ is the pion's four-momentum, $P^2=-m_\pi^2$ and $n\cdot P = -m_\pi$, with $m_\pi$ being the pion's mass; and $\zeta$ is the scale at which the dressed-quark propagators and pion Bethe-Salpeter amplitude are computed.

How should one expect the leading-twist PDA defined by Eq.\,\eqref{pionPDA} to behave?  In order to answer this question, recall that the pion multiplet contains a charge-conjugation eigenstate: $\pi^0$, the neutral pion.  Therefore, the peak in the leading Chebyshev moment [$m=0$ in Eq.\,\eqref{Chebyshev}] of each of the three significant scalar functions that appear in the expression for $\Gamma_\pi(q;P)$ occurs at $q_{+-} = 0$, \emph{i.e}.\ at zero relative momentum \cite{Maris:1997tm, Maris:1999nt, Qin:2011xq}.  Moreover, these Chebyshev moments are monotonically decreasing with $q_{+-}^2$.  It follows that $\varphi_\pi(x;\zeta^2)$ should exhibit a single maximum, which appears at $x=1/2$, \emph{viz}.\ $\forall \zeta^2>0$, $\varphi_\pi(x;\zeta^2)$ is a symmetric, concave function on $x\in [0,1]$.

Nonperturbative approaches to problems in hadron physics that possess a traceable connection to QCD are typically formulated in Euclidean space (see, \emph{e.g}.\ Sec.\,1.3 of Ref.\,\cite{Roberts:2012sv}).  In connection with the PDA, this might seem to be a problem because the light-front is a concept peculiar to Minkowski space.  The difficulty can be overcome by appealing to the fact that the PDA in Eq.\,\eqref{pionPDA} is completely characterised by its moments
\begin{equation}
\langle x^m \rangle = \int_0^1 \! dx\, x^m\, \varphi_\pi(x) \quad \mbox{or} \quad
\langle x_\Delta^m \rangle = \int_0^1 \! dx\, (2 x-1)^m\, \varphi_\pi(x) \,,
\end{equation}
which may be obtained via
\begin{equation}
f_\pi (n\cdot P)^{m+1} \langle x^m \rangle =
{\rm tr}_{\rm CD} Z_2 \! \int_{dq}^\Lambda \!\! (n\cdot q_+)^m\, \gamma_5\gamma\cdot n\, \chi_\pi(q;P)\,;
\end{equation}
and recognising that such moments can be computed from information produced by Euclidean space analyses of the pion bound-state problem.  A novel technique for achieving this outcome was introduced in Ref.\,\cite{Chang:2013pqS}; and, formulated in the continuum, the procedure enables one to compute arbitrarily many moments and hence accurately reconstruct the pion's PDA.

%
%

\begin{figure}[t]
\centerline{\includegraphics[clip,width=0.6\textwidth]{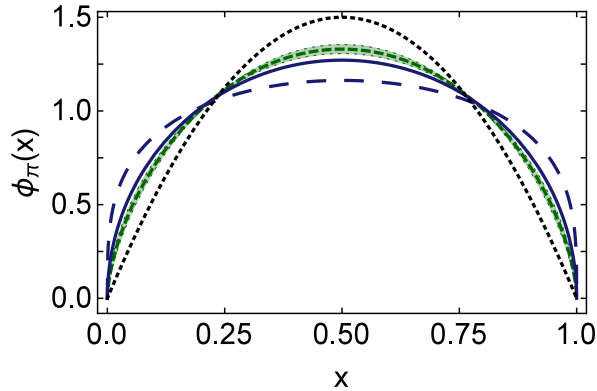}}

\caption{\label{figPDAlQCD} Predictions for the pion's twist-two valence-quark PDA, all computed at a resolving scale $\zeta=2\,$GeV$=:\zeta_2$: \emph{Solid curve} (blue) -- DB truncation \cite{Chang:2013pqS}; long-dashed curve (blue) -- RL truncation \cite{Chang:2013pqS}; and short-dashed curve (green) within (green) shaded area -- lQCD \cite{Braun:2015axa}, as determined following the method in Ref.\,\cite{Cloet:2013ttaS}.  In connection with the lattice curve, the band reflects the error quoted in Eq.\,\eqref{xDelta}.  The dotted curve (black) is the conformal limit result in Eq.\,\eqref{PDAcl}.}
\end{figure}

The RL- and DB-truncation pion PDAs computed in Ref.\,\cite{Chang:2013pqS} at a resolving scale $\zeta = 2\,$GeV$=:\zeta_2$ are depicted in Fig.\,\ref{figPDAlQCD}.  They yield
\begin{equation}
\label{xDeltaDBRL}
\langle x_\Delta^2 \rangle_{\rm DB} = 0.25\,,\quad
\langle x_\Delta^2 \rangle_{\rm RL} = 0.28\,.
\end{equation}
The figure also displays the conformal limit result in Eq.\,\eqref{PDAcl}, for which $\langle x_\Delta^2 \rangle =0.2$, and a pointwise form inferred, using the method in Ref.\,\cite{Cloet:2013ttaS}, from a lQCD computation of the PDAs first non-trivial moment \cite{Braun:2015axa}:
\begin{equation}
\label{xDelta}
\langle x_\Delta^2 \rangle_{\rm lQCD} = 0.236 \pm 0.006\,.
\end{equation}
(The error quoted here accounts for uncertainties in the chiral extrapolation and nonperturbatively determined renormalisation factors; but does not include effects associated with lattice spacing.)

It is apparent from Fig.\,\ref{figPDAlQCD} that all calculations predict a PDA that is significantly broader than the conformal limit result in Eq.\,\eqref{PDAcl}.  This important outcome is a direct expression of DCSB, as we now explain.  The continuum predictions were computed at a low renormalisation scale in the chiral limit, whereat the quark mass function owes entirely to DCSB; and, on the domain $p^2\in[0,\zeta^2]$, the nonperturbative interactions responsible for DCSB produce significant structure in the dressed-quark's self-energy, Fig.\,\ref{gluoncloud}.  The PDA is an integral of the pion's Bethe-Salpeter wave function, whose pointwise behaviour is rigorously connected with that of the quark self-energy [see Eqs.\,\eqref{gGTrelations}].  Hence, the structure of the pion's distribution amplitude at the hadronic scale is a pure expression of DCSB.  As the scale is removed to extremely large values, phase space growth diminishes the impact of nonperturbative DCSB interactions, so that the PDA relaxes to its asymptotic form.

It is appropriate here to reflect upon and explain the pointwise difference between the DB and RL results in Fig.\,\ref{figPDAlQCD}.  Note first that low-$m$ moments are sensitive to the behaviour of $\varphi_\pi(x)$ in the neighbourhood of $x=1/2$, whereas high-$m$ moments are sensitive to its endpoint behaviour.
Then consider that RL-kernels ignore DCSB in the quark-gluon vertex.  Therefore, to describe a given body of phenomena, they must shift all DCSB-strength into the infrared behaviour of the quark propagator, whilst nevertheless maintaining perturbative behaviour for $p^2>\zeta^2$.  This requires $B(p^2)$ to be large at $p^2=0$, but drop quickly, behaviour which influences $\varphi_\pi(x)$ via Eq.\,\eqref{BGTrelation}.  The concentration of strength at $p^2\simeq 0$ forces large values for the small-$m$ moments [see Eq.\,\eqref{xDeltaDBRL}], which translates into a broad distribution.
In contrast, the DB-kernel builds DCSB into the quark-gluon vertex and its impact is therefore shared between more elements of a calculation.  Hence a smaller value of $B(p^2=0)$ is capable of describing the same body of phenomena; and this self-energy need fall less rapidly in order to reach the common asymptotic limit.  It follows that the low-$m$ moments are smaller and the distribution is narrower.

The near match between the DSE-DB prediction and the lQCD curve is also significant.  An earlier lQCD result \cite{Braun:2006dg} produced a PDA that was in better agreement with the DSE-RL curve; but, as anticipated in Ref.\,\cite{Cloet:2013ttaS}, improvements in the lattice simulations have produced a PDA in good agreement with the DSE-DB prediction, which is the more realistic result, as explained in Sec.\,\ref{secCBSP}.  Contemporary theory has therefore converged on a pointwise form for the pion's twist-two valence-quark PDA; and that form is \cite{Mikhailov:1986be, Petrov:1998kg, Brodsky:2006uqa, Chang:2013pqS, Cloet:2013ttaS}:
\begin{equation}
\label{phiH}
\varphi_\pi(x;\zeta_2^2) \approx \frac{8}{\pi} \sqrt{x (1-x)}\,.
\end{equation}
It follows that $\varphi_\pi^{\rm asy}(x)$ is a poor approximation to $\varphi_\pi(x;Q^2)$ at all momentum-transfer scales that are either now accessible to experiments involving pion elastic or transition processes, or will become so in the foreseeable future \cite{Volmer:2000ek, Horn:2006tm, Tadevosyan:2007yd, Huber:2008id, Uehara:2012ag, Holt:2012gg, Dudek:2012vr}.  Available information indicates that the pion's PDA is significantly broader at these scales; and hence that predictions of leading-order, leading-twist formulae involving $\varphi^{\rm asy}_\pi(x)$, such as Eqs.\,\eqref{pionUV} and \eqref{BLuv}, must be a misleading guide to interpreting and understanding contemporary experiments.  At accessible energy scales a better guide is obtained by using the dilated PDAs depicted in Fig.\,\ref{figPDAlQCD} in such formulae, as we will now proceed to illustrate.

\section{Elastic Electromagnetic Form Factors}
\label{SecElastic}
\subsection{Pion}
\label{SecElasticPion}
The first publication by the JLab $F_\pi$ Collaboration \cite{Volmer:2000ek} signalled the beginning of a new era in probing the pion's internal structure.  Subsequent measurements \cite{Horn:2006tm, Tadevosyan:2007yd, Horn:2007ug, Huber:2008id, Blok:2008jy} confirmed the observed data trend and this led to a widespread perception that, with a momentum transfer of $Q^2=2.45\,$GeV$^2$, one is still far from the resolution region wherein the pion behaves like a simple quark-antiquark pair, \emph{i.e}.\ far from establishing validity of Eq.\,\eqref{pionUV}.  This conclusion was based on the assumption that Eq.\,\eqref{PDAcl} is valid at $Q^2=2.45\,$GeV$^2$ and hence Eq.\,\eqref{Fpicl} provides the appropriate pQCD prediction with which to compare, in which case
\begin{equation}
\label{pionUV4}
Q^2 F_\pi(Q^2) \stackrel{Q^2=4\,{\rm GeV}^2}{\approx} 0.15\,,
\end{equation}
The result in Eq.\,\eqref{pionUV4} is a factor of $2.7$ smaller than the empirical value quoted at $Q^2 =2.45\,$GeV$^2$ \cite{Huber:2008id}: $0.41^{+0.04}_{-0.03}$; and a factor of three smaller than that computed at $Q^2 =4\,$GeV$^2$ in Ref.\,\cite{Maris:2000sk}.  At the time, Ref.\,\cite{Maris:2000sk} provided the only prediction for the pointwise behaviour of $F_\pi(Q^2)$ that was both applicable on the entire spacelike domain then mapped reliably by experiment and confirmed thereby.

In this case the perception of a mismatch and a real discrepancy are not equivalent because, as we have just elucidated, one can argue that $Q^2=4\,$GeV$^2$ is not within the domain $\Lambda_{\rm QCD}^2/Q^2\simeq 0$ upon which Eq.\,\eqref{PDAcl} is valid.  This being so, and given the successful prediction in Ref.\,\cite{Maris:2000sk}, one is naturally led to ask whether the methods used therein can address the issue of the ultimate validity of Eq.\,\eqref{pionUV}.
Until recently, the answer was ``no'', owing to an over-reliance on brute numerical methods in continuum bound-state calculations.  That changed, however, with an appreciation of the versatility inherent in the refinement of known techniques \cite{Nakanishi:1963zz, Nakanishi:1969ph, Nakanishi:1971} which enabled the computation of the pion's light-front wave function reviewed in Sec.\,\ref{secLTPDA}.  Those methods also enable a computation of the pion's electromagnetic form factor to arbitrarily large-$Q^2$ and the correlation of that result with Eq.\,\eqref{pionUV} using the consistently computed distribution amplitude, $\varphi_\pi(x)$.  This effort is detailed in Ref.\,\cite{Chang:2013niaS}; and we will summarise it here.

\begin{figure}[t]
\begin{minipage}[t]{1.0\textwidth}
\begin{minipage}{0.40\textwidth}
\centerline{\includegraphics[clip,width=0.97\textwidth]{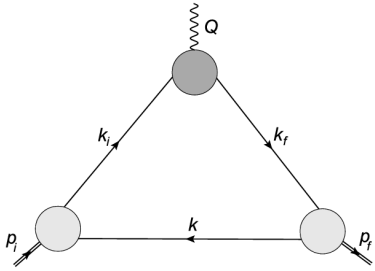}}
\end{minipage}
\begin{minipage}{0.59\textwidth}
\rightline{\parbox{28em}{\caption{\label{FpiVertex}
Pictorial representation of Eq.\,\eqref{RLFpi}, which defines the RL truncation of the pion elastic form factor.  The unamputated photon quark vertex $\chi_\mu(k_f,k_i) = S(k_f) \Gamma_\mu(k_f,k_i) S(k_i)$, where $\Gamma_\mu(k_f,k_i)$ is the unamputated vertex, which is indicated by the dark-shaded vertex.  The pion Bethe-Salpeter amplitudes are marked by the light-shaded vertices and the solid internal lines are the dressed-quark propagators.
}}}
\end{minipage}
\end{minipage}
\end{figure}

At leading-order in the DSE truncation scheme explained in Refs.\,\cite{Munczek:1994zz, Bender:1996bb, Binosi:2016rxzd}, i.e.\ in RL truncation, the pion form factor is given by the diagram in Fig.\,\ref{FpiVertex}, which corresponds to the following expression:
\begin{eqnarray}
\nonumber
K_\mu F_\pi(Q^2) & = & N_c {\rm tr}_{\rm D} 
\int\! \frac{d^4 k}{(2\pi)^4}\,
\chi_\mu(k+p_f,k+p_i) \\
&& \times \Gamma_\pi(k_i;p_i)\,S(k)\,\Gamma_\pi(k_f;-p_f)\,, \quad\label{RLFpi}
\end{eqnarray}
where $Q$ is the incoming photon momentum, $p_{f,i} = K\pm Q/2$, $k_{f,i}=k+p_{f,i}/2$, and the remaining trace is over spinor indices.  Isospin symmetry is assumed in Eq.\,\eqref{RLFpi}, so that $S_u=S=S_d$; and $\chi_\mu(k_f,k_i)$ is the unamputated dressed-quark-photon vertex, which should also be computed in rainbow-ladder truncation.
It is worth remarking that the dominant effect of corrections to RL truncation is a modification of the power associated with the logarithmic running in Eq.\,\eqref{pionUV}; but since that running is slow, the diagrams omitted have no material impact on the main course of this discussion.  Notwithstanding that, it is possible to obtain a reliable estimate of their impact following the method described in connection with Fig.\,\ref{figDSEprediction} in Sec.\,\ref{SecTransition}.

The leading-order prediction for the pion form factor is determined once an interaction kernel is specified for the rainbow gap equation.  The authors of Ref.\,\cite{Chang:2013niaS} used the same form employed in Ref.\,\cite{Chang:2013pqS} to produce the long-dashed (blue) curve in Fig.\,\ref{figPDAlQCD} and hence the only new element in the computation of $F_\pi(Q^2)$ in Ref.\,\cite{Chang:2013niaS} was $\chi_\mu(k_f,k_i)$.  Instead of solving a Bethe-Salpeter equation for that vertex, Ref.\,\cite{Chang:2013niaS} used an \emph{Ansatz}, which expedited completion of the $F_\pi(Q^2)$  computation.  %
That is a valid strategy so long as nothing essential to understanding the form factor is lost thereby.  This was established by noting that since the \emph{Ansatz} is obtained using the gauge technique \cite{Delbourgo:1977jc}, the vertex satisfies the longitudinal vector WGTI \cite{Ward:1950xp, Green:1953te, Takahashi:1957xn}, is free of kinematic singularities, reduces to the bare vertex in the free-field limit, and has the same Poincar\'e transformation properties as the bare vertex.  Moreover, numerical solutions of the RL Bethe-Salpeter equation for the vertex \cite{Maris:1999bh} and algebraic analyses of vertex structure \cite{Chang:2010hb, Bashir:2011dp, Qin:2013mtaS} show that nonperturbative corrections to the bare vertex are negligible for spacelike momenta $Q^2\gtrsim 1\,$GeV$^2$.
A deficiency of the \emph{Ansatz} is omission of explicit nonanalytic structures associated with the $\rho$-meson pole.  However, such features only have an impact on $Q^2 r_{\pi}^2 \lesssim 1$, where $r_\pi$ is the charged pion's electromagnetic radius, and are otherwise immaterial at spacelike momenta \cite{Alkofer:1993gu, Roberts:1994hh, Roberts:2000aa}.  Furthermore, salient aspects thereof are included implicitly, \emph{e.g}.\ their influence on pion radii, as explained in connection with Eqs.\,(2.3.39), (2.3.40) in Ref.\,\cite{Roberts:2000aa}.

The electromagnetic pion form factor, computed from Eq.\,\eqref{RLFpi} using the elements and procedures described above, is depicted as the solid curve in Fig.\,\ref{Fig1CLpion} and marked as curve-A in the lower panel.
It is evident from the upper panel that this prediction is practically indistinguishable from that described in Ref.\,\cite{Maris:2000sk} on the spacelike domain $Q^2 < 4\,$GeV$^2$, which was the largest value computable reliably in that study.  Critically, however, the new prediction extended to arbitrarily large momentum transfers: owing to the improved algorithms, it describes an unambiguous continuation of the earlier DSE prediction to the entire spacelike domain and thereby achieved a longstanding goal.
For comparison, the current status of lattice-QCD calculations of $F_\pi(Q^2)$ is described in Refs.\,\protect\cite{Alexandrou:2011iu, Renner:2012yh}.  Results with quantitatively controlled uncertainties are beginning to become available.  Within errors, the estimated charge radius matches experiment; and simulations are also exploring a non-zero but still low $Q^2$ domain ($0<Q^2<1\,$GeV$^2$).  Various systematic uncertainties become more challenging with increasing $Q^2$, making access to a larger domain difficult at present \cite{Brandt:2013ffb}.

The momentum reach of the improved continuum techniques for computation of form factors is emphasised by the lower panel in Fig.\,\ref{Fig1CLpion}.  The prediction for $F_\pi(Q^2)$ is depicted on the domain $Q^2\in [0,20]\,$GeV$^2$ but was computed in Ref.\,\cite{Chang:2013niaS} out to $Q^2=100\,$GeV$^2$.  If necessary, reliable results could readily have been obtained at even higher values.  That is not required, however, because the longstanding questions revolving around $F_\pi(Q^2)$, which we reiterated in opening this subsection, may be answered via Fig.\,\ref{Fig1CLpion}.  In this connection, a key feature of the prediction for $Q^2 F_\pi(Q^2)$ is the maximum at $Q^2\approx 6\,$GeV$^2$ that is evident in the right panel of  Fig.\,\ref{Fig1CLpion}.  The domain upon which the flattening of the curve associated with this extremum is predicted to occur will be accessible to next-generation experiments \cite{E1206101, E1207105}; and, importantly, if Ref.\,\cite{E1207105} achieves its full potential, then it will be possible to distinguish between the theoretical prediction and the monopole fitted to data in Ref.\,\cite{Amendolia:1986wj}.

A quick glance at the lower panel of Fig.\,\ref{Fig1CLpion} suggests that a maximum is necessary if $Q^2 F_\pi(Q^2)$ is ever to approach the value predicted by pQCD, Eq.\,\eqref{pionUV}.  In this connection, too, Ref.\,\cite{Chang:2013niaS} had something to add.  The result in Eq.\,\eqref{pionUV4} is associated with curve-D in the right panel of Fig.\,\ref{Fig1CLpion}, which is typically plotted in such figures and described as the prediction of pQCD.  That would be true if, and only if, the pion's valence-quark distribution amplitude were well described by $\varphi^{\rm cl}(x)$ in Eq.\,\eqref{PDAcl} at the scale $Q^2\sim 4\,$GeV$^2$.  However, that is not the case, as we saw in Sec.\,\ref{secLTPDA}.

The sensible comparison with pQCD should be drawn as follows.  Using precisely the interaction that was employed to compute $F_\pi(Q^2)$, one obtains the RL truncation result described in Sec.\,\ref{secLTPDA}, \emph{viz}.
\begin{equation}
\label{phiRLreal}
\varphi_\pi(x;\zeta_2^2) \approx 1.74 \, [x (1-x)]^{0.29} \,.
\end{equation}
This amplitude provides a far better choice than $\varphi^{\rm cl}(x)$ when calculating the pQCD prediction appropriate for comparison with contemporary experiments.  That computed result is drawn as curve-C in the lower panel of Fig.\,\ref{Fig1CLpion}, \emph{i.e}.\ this curve is the pQCD prediction obtained when Eq.\,\eqref{phiRLreal} is used in Eqs.\,\eqref{pionUV}--\eqref{wphi}.

Stated simply, curve-C in the lower panel of Fig.\,\ref{Fig1CLpion} is the pQCD prediction obtained when the pion valence-quark PDA has a form appropriate to the scale accessible in modern experiments.  Its magnitude is markedly different from that obtained using the conformal-limit PDA in Eq.\,\eqref{PDAcl}, \emph{viz}.\ curve-D, which is only valid at truly asymptotic momenta.
The meaning of ``truly asymptotic'' is readily illustrated.  The PDA in Eq.\,\eqref{phiRLreal} produces $\mathpzc{w}_\varphi^2 = 3.3$, which is to be compared with the value computed using the conformal-limit PDA: $\mathpzc{w}^{\rm asy}_\varphi = 1.0$.  Applying leading-order ERBL evolution \cite{Lepage:1979zb, Efremov:1979qk, Lepage:1980fj} to the PDA in Eq.\,\eqref{phiRLreal}, one must reach momentum transfer scales $Q^2 > 1000\,$GeV$^2$ before $\mathpzc{w}_\varphi^2 < 1.6$, \emph{i.e}.\ before $\mathpzc{w}_\varphi^2$ falls below half its original value, because ERBL evolution is logarithmic.

Given these observations, the near agreement between the pertinent perturbative QCD prediction in Fig.\,\ref{Fig1CLpion} (lower panel, curve-C) and the continuum prediction for $Q^2 F_\pi(Q^2)$ (lower panel, curve-A) is striking.  It highlights that a single interaction kernel for the continuum bound-state problem has completed the task of unifying the pion's electromagnetic form factor and its valence-quark distribution amplitude.

Moreover, this leading-order, leading-twist QCD prediction, obtained with a pion valence-quark PDA evaluated at a scale appropriate to the experiment, Eq.\,\eqref{phiRLreal}, underestimates the full continuum computation by merely an approximately uniform 15\% on the domain depicted.
The small mismatch is explained by a combination of higher-order, higher-twist corrections to Eq.\,\eqref{pionUV} in pQCD on the one hand, and shortcomings in the RL truncation, which predicts the correct power-law behaviour for the form factor but not precisely the right anomalous dimension in the strong coupling calculation on the other hand.
Hence, as anticipated earlier \cite{Maris:1998hc} and expressing a result that can be understood via the behaviour of the dressed-quark mass-function in Fig.\,\ref{gluoncloud}, one should expect dominance of hard contributions to the pion form factor for $Q^2\gtrsim 8\,$GeV$^2$.  Expressed differently, on $Q^2\gtrsim 8\,$GeV$^2$, it is predicted that $F_\pi(Q^2)$ will exhibit precisely the momentum-dependence anticipated from QCD, the power-law behaviour plus logarithmic violations of scaling, but with the normalisation fixed by a pion wave function whose dilation with respect to $\varphi^{\rm cl}(x)$ is a definitive signature of DCSB, which is a crucial feature of the Standard Model.

Efforts to improve upon this calculation by using more sophisticated truncations, such as the DB scheme, are underway.  At present it seems that the most promising route to success in this endeavour lies in calculating the pion's light-front wave function, using techniques similar to those that delivered the pion's leading-twist PDA; and therewith computing the elastic form factor via an overlap representation \cite{Drell:1969km, West:1970av}.  In this way, the pion form factor becomes a byproduct of calculating the pion's generalised parton distribution, as illustrated in Ref.\,\cite{Mezrag:2014jkaS}.

\subsection{Kaon}
It is worth commenting briefly here on the kaon's electromagnetic form factor, $F_K(Q^2)$.  As noted in Sec.\,\ref{sec:pion-ff-exp}, scattering of high-energy charged kaons from atomic electrons has delivered direct measurements of $F_K(Q^2)$ out to $Q^2$=0.10 GeV$^2$ \cite{Dally:1980dj, Amendolia:1986ui}.  This data is displayed in the upper panel of Fig.\,\ref{fig:fk_exp_current} along with the prediction in Ref.\,\cite{Maris:2000sk}.  That study provided a unified description of pion and kaon elastic form factors on $Q^2\in[0,4]\,$GeV$^2$; but, owing to limitations in the numerical methods employed, it was unable to produce a reliable result on $Q^2>4\,$GeV$^2$, something evident in the lower panel of Fig.\,\ref{fig:fk_exp_current}.  The methods employed in Ref.\,\cite{Chang:2013niaS} can surmount this difficulty and the required analysis is underway.  However, it is not yet complete.  Herein, therefore, we consider what might be learnt from the relevant hard-scattering formula \cite{Lepage:1980fj}, a generalisation of Eq.\,\eqref{pionUV}:
\begin{equation}
\exists \bar Q_0 > \Lambda_{\rm QCD} \; | \;
Q^2 F_K(Q^2) \stackrel{Q^2>\bar Q_{0}^2}{\approx} 16 \pi \alpha_s(Q^2) f_K^2 \mathpzc{w}_K^2(Q^2)\,,
\label{kaonUV}
\end{equation}
with \cite{Agashe:2014kda} $f_K=0.110\,$GeV and, for the $K^+$:
\begin{subequations}
\begin{eqnarray}
\mathpzc{w}_K^2 & =& e_{\bar s} \mathpzc{w}_{\bar s}^2 + e_{u}\mathpzc{w}_u^2,  \\
\label{weightings}
\mathpzc{w}_{\bar s} & = & \frac{1}{3}\int_0^1 dx\, \frac{1}{1-x}\,\varphi_K(x) \,, \quad
\mathpzc{w}_u  = \frac{1}{3}\int_0^1 dx\, \frac{1}{x}\,\varphi_K(x) \,,
\end{eqnarray}
\end{subequations}
where $e_{u}=2  e_{\bar s} = (2/3)$ and $\varphi_K(x)$ is the kaon's twist-two PDA.

\begin{figure}[t]
\centerline{\includegraphics[clip,width=0.6\textwidth]{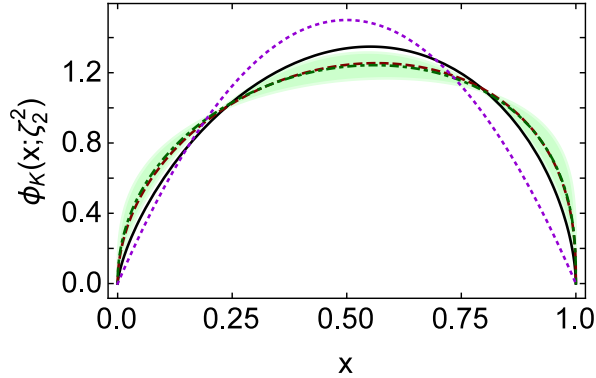}}

\caption{\label{figPDAK} Predictions for the kaon's twist-two valence-quark PDA,  computed at a resolving scale $\zeta=2\,$GeV$=:\zeta_2$: \emph{Solid curve} (black) -- DB truncation \cite{Shi:2014uwaS, Shi:2015esaS}; dot-dashed (green) -- PDA inferred from lQCD moments in Ref.\,\cite{Arthur:2010xf} using the method described in Refs.\,\cite{Cloet:2013ttaS, Segovia:2013ecaS}.  The dashed (red) curve is the PDA discussed in connection with Eq.\,\eqref{PDAKDSEmod}; and
the dotted (violet) curve is the conformal limit result in Eq.\,\eqref{PDAcl}.}
\end{figure}

The large-$Q^2$ behaviour is plainly determined by the shape of $\varphi_K(x)$, information about which is today available from DSE- and lattice-QCD analyses.  The most sophisticated DSE analyses produce a PDA that may be represented as \cite{Shi:2014uwaS, Shi:2015esaS}:
\begin{equation}
\label{DSEkaon}
\varphi_K^{\rm DSE}(x;\zeta_2^2) = 3.29 \, x^{0.71} (1-x)^{0.58}\,,
\end{equation}
which is a distribution whose peak lies at $x>(1/2)$, \emph{i.e}.\ is skewed to indicate that the heavier $\bar s$-quark in the $K^+$ is carrying more of the bound-state's momentum (see Fig.\,\ref{figPDAK}).  This PDA yields the first row in the following array:
\begin{equation}
\label{kaonmoments}
\begin{array}{l|cl}
    & \langle (2x-1) \rangle & \langle (2x-1)^2 \rangle \\\hline
\mbox{DSE\,\cite{Shi:2014uwaS, Shi:2015esaS}} & 0.040\phantom{(2)} & 0.23\phantom{\times 1.1} \\
\mbox{DSE}_{\rm mod} & 0.040\phantom{(2)} & 0.23 \times 1.1 \\
\mbox{lQCD\,\cite{Arthur:2010xf}} & 0.036(2) & 0.26(2)
\end{array}\,.
\end{equation}
The third row in Eq.\,\eqref{kaonmoments} reports the most recent information available from lQCD.  Analysed according to the method in Refs.\,\cite{Cloet:2013ttaS, Segovia:2013ecaS}, it corresponds to
\begin{equation}
\label{lQCDkaon}
\varphi_K^{\rm lQCD}(x;\zeta_2^2) = \mathpzc{n}_{\alpha\beta} \, x^\alpha (1-x)^\beta,\,\quad
\alpha = 0.48 \pm 0.15\,, \; \beta = 0.38 \pm 0.15\,,
\end{equation}
where $\mathpzc{n}_{\alpha\beta} = 1/B(1+\alpha,1+\beta)$.

%
%
%

The moments in rows~1 and 3 of Eq.\,\eqref{kaonmoments} do not appear too different; and the corresponding PDAs agree, within errors, as highlighted by Fig.\,\ref{figPDAK}, although the lQCD result exhibits greater dilation.  Owing to the weighting factors in Eqs.\,\eqref{weightings}, however, the seemingly small differences between these PDAs have a large impact.  This is apparent in the lower panel of Fig.\,\ref{fig:fk_exp_current}: the long-dashed (black) and dot-dashed (green) curves display the predictions of Eq.\,\eqref{kaonUV} obtained with Eqs.\,\eqref{DSEkaon} and \eqref{lQCDkaon}, respectively; and the green band highlights the uncertainty introduced by the errors on $\alpha$, $\beta$ in Eq.\,\eqref{lQCDkaon}.  Plainly, a reliable prediction for the behaviour of $Q^2 F_K(Q^2)$ at potentially accessible momenta using Eq.\,\eqref{kaonUV} depends very sensitively on the amount of distortion and/or dilation in the kaon's PDA.  This fact materially increases the importance of extending the direct calculation of $F_\pi(Q^2)$ in Ref.\,\cite{Chang:2013niaS} to $F_K(Q^2)$.

In order to elucidate further, we considered the impact of supposing that the DSE predictions for the moments in Eq.\,\eqref{kaonmoments} are only accurate to 10\%, \emph{i.e}.\ we considered the effects of the replacements $\{0.040,0.23\} \to \{0.040\pm 10\%,0.23\pm 10\%\}$.  One finds in this way that the variation in $\langle (2x-1)\rangle$ is immaterial; but the small change in $\langle (2x-1)^2\rangle$ has a significant influence.  Row~2 in Eq.\,\eqref{kaonmoments} expresses a 10\% increase in $\langle (2x-1)^2\rangle$; and this pair of moments yields the following PDA:
\begin{equation}
\label{PDAKDSEmod}
\varphi_K^{\rm DSE_{\rm mod}}(x;\zeta_2^2) = 2.33\, x^{0.51} (1-x)^{0.39}\,,
\end{equation}
which, as evident in Fig.\,\ref{figPDAK}, is noticeably more dilated than the original PDA and, in fact, almost indistinguishable from the central lQCD result: the curves lie almost atop one another; and their predictions for $Q^2 F_K(Q^2)$ are equivalent (see Fig.\,\ref{fig:fk_exp_current}).  Plainly, data anticipated from JLab\,12 \cite{E12-09-011} can potentially contribute a great deal to forming an accurate picture of the kaon's internal structure.

Since Eqs.\,\eqref{pionUV}, \eqref{kaonUV} should also be valid for large timelike-$Q^2=(-t)$, it is worth considering the prediction they make for the ratio $F_K(t)/F_\pi(t)$, which has been measured  in $e^+ e^-$ annihilation on a large domain, with an upper bound of $s_U=17.4\,$GeV$^2$ \cite{Seth:2012nnS}: $|F_K(s_U)|/|F_\pi(s_U)| = 0.92(5)$.  According to Eqs.\,\eqref{pionUV}, \eqref{kaonUV}:
\begin{equation}
\frac{F_K(t)}{F_\pi(t)} = \frac{f_K^2}{f_\pi^2} \frac{\mathpzc{w}_K^2(t)}{\mathpzc{w}_\pi^2(t)}\,.
\end{equation}
Plainly, on $\Lambda_{\rm QCD}/|t| \simeq 0$, this ratio has the value $f_K^2/f_\pi^2=1.43$; a limit which Eqs.\,\eqref{pionUV}, \eqref{kaonUV} predict is, at least initially, approached from below as $t$ grows from zero because the ratio is unity at $t=0$.  Working with the most sophisticated DSE results available, Eqs.\,\eqref{phiH}, \eqref{DSEkaon}, and using leading-order ERBL evolution, the result at $t=s_U$ is:
\begin{equation}
\frac{F_K(s_U)}{F_\pi(s_U)} = 1.16\,.
\label{ratioFKfpi}
\end{equation}
If the PDAs in Figs.\,\ref{figPDAlQCD}, \ref{figPDAK} that were inferred from lQCD results are used to compute this ratio, the midpoint value is 50\% greater, owing largely to the increased dilation of the kaon's PDA.

There is some tension between the published empirical value \cite{Seth:2012nnS} and Eq.\,\eqref{ratioFKfpi}, which differ by $\lesssim 5$-standard-deviations.  Whilst this is far less than the 9-standard-deviation discrepancy with pQCD reported in Ref.\,\cite{Seth:2013eaa}, which arises if one (misguidedly) uses the conformal limit result for both meson PDAs [Eq.\,\eqref{PDAcl}] in order to evaluate the ratio, the mismatch is still cause for further consideration.  The DSE result for the kaon PDA in Eq.\,\eqref{DSEkaon} only exhibits modest skewing and dilation; but these distortions must be decreased even more, so that $\varphi_K(x)$ quite closely resembles $\varphi^{\rm cl}(x)$, without affecting $\varphi_\pi(x)$, if the value of the ratio in Eq.\,\eqref{ratioFKfpi} is to be reduced.  In that case, however, the long-dashed curve in the lower panel of Fig.\,\ref{fig:fk_exp_current} shifts even further away from the prediction in Ref.\,\cite{Maris:2000sk}, \emph{viz}.\ toward the dotted (blue) curve.
It is possible that the normalisation of the $F_{K,\pi}(s_U)$ measurements should be reconsidered \cite{Holt:2012gg}.  There is certainly a puzzle, which can be highlighted by comparing the value of $F_\pi(Q^2=|s_U|)=0.42/|s_U|$ computed in Ref.\,\cite{Chang:2013nia} and $|F_\pi(s_U)|=0.84(5)/s_U$ reported in Ref.\,\cite{Seth:2012nnS}.  The computation in Ref.\,\cite{Chang:2013nia} agrees with all available, reliable spacelike data, and the calculated value of $F_\pi(|s_U|)$ is a factor of four larger than the result obtained from Eq.\,\eqref{pionUV} using $\varphi^{\rm cl}(x)$.  It is nevertheless still a factor of two smaller than the reported timelike experimental value.
Internally consistent calculations, capable of generating predictions which connect the spacelike and timelike large-$|Q^2|$  behaviour of $F_{K,\pi}$, would be useful in answering this question; but reaching that goal is a challenging task \cite{Bruch:2004py, deMelo:2005cy, Brodsky:2007hb, Cloet:2008fw, Choi:2008yj, Bulava:2015qjz}.

\section{Electromagnetic Transition Form Factor}
\label{SecTransition}
The neutral pion electromagnetic transition form factor, $G_{\gamma^\ast\gamma \pi^0}(Q^2)$, is a very particular expression of this meson's internal structure.  It is measured in the process $\gamma^\ast_Q \gamma\to \pi^0$, which is fascinating because its complete understanding demands a framework capable of simultaneously combining an explanation of the essentially nonperturbative Abelian anomaly \cite{Adler:1969gk, Bell:1969ts, Adler:2004ih}, which determines $G_{\gamma^\ast\gamma \pi^0}(Q^2\simeq 0)$, with the features of perturbative QCD that govern the behaviour of $G_{\gamma^\ast\gamma \pi^0}(Q^2)$ on the domain of ultraviolet momenta, Eq.\,\eqref{BLuv}.

In the chiral limit, the Abelian anomaly entails that
\begin{equation}
\label{Ggpg0}
2 f_{\pi}^0 G_{\gamma^\ast\gamma \pi^0}(Q^2=0) = 1\,,
\end{equation}
where $f_\pi^0\approx 0.09\,$GeV is the chiral-limit value of the charged pion's leptonic decay constant; and thereby locks the rate of this transition to the strength of DCSB in the Standard Model.  Corrections to Eq.\,\eqref{Ggpg0}, arising from nonzero and unequal light-quark masses, have been computed \cite{Bernstein:2011bx}: they are small; but, curiously, extant measurements suggest that the calculations actually overestimate their size \cite{Larin:2010kq}.

At the other extreme, as we have already seen, the property of factorisation in QCD hard scattering processes leads to the inviolable prediction in Eq.\,\eqref{BLuv}; and at this point it is natural to compare Eq.\,\eqref{BLuv} with Eq.\,\eqref{pionUV}, the analogue for $F_\pi(Q^2)$.  With both normalised to unity at $Q^2=0$, then on any momentum domain for which the asymptotic limit of both is valid, the transition form factor is $\pi/[2\alpha_s(Q^2)]$-times \emph{larger}: at $Q^2=4\,$GeV$^2$, this is a factor of \emph{five}.

The prediction in Eq.\,\eqref{BLuv} and the manner by which it is approached are currently receiving keen scrutiny (\emph{e.g}.\ Refs.\,\cite{Radyushkin:2009zg, Agaev:2010aq, Roberts:2010rnS, Brodsky:2011yv, Bakulev:2011rp, Brodsky:2011xx, Bakulev:2012nh, ElBennich:2012ij, Lucha:2013yca, Dorokhov:2013xpa}) following publication of data by the BaBar Collaboration \cite{Aubert:2009mc}.  Whilst those data agree with earlier experiments on their common domain of momentum-transfer \cite{Behrend:1990sr, Gronberg:1997fj}, they are unexpectedly far \emph{above} the prediction in Eq.\,\eqref{BLuv} on $Q^2\gtrsim 10\,$GeV$^2$.  Numerous authors have attempted to reconcile the BaBar measurements with Eq.\,\eqref{BLuv}, typically producing a transition form factor whose magnitude on $Q^2\gtrsim 10\,$GeV$^2$ exceeds the ultraviolet limit, without explaining how that limit might finally be recovered \cite{Radyushkin:2009zg, Agaev:2010aq, Dorokhov:2013xpa} or how their results might be reconciled with modern measurements of $F_\pi(Q^2)$ \cite{Volmer:2000ek, Horn:2006tm, Tadevosyan:2007yd, Huber:2008id}.  Others, however, argue that the BaBar data is not an accurate measure of the transition form factor \cite{Roberts:2010rnS, Brodsky:2011yv, Bakulev:2011rp, Brodsky:2011xx, Bakulev:2012nh, ElBennich:2012ij, Lucha:2013yca}.  Significantly, data subsequently published by the Belle Collaboration \cite{Uehara:2012ag} appear to be in general agreement with Eq.\,\eqref{BLuv}.

One can argue \cite{Raya:2015gvaS} that the limit in Eq.\,\eqref{BLuv} should either be approached from below or only exceeded marginally, with logarithmic approach to the ultraviolet limit from above in that event.  It is worth recapitulating the reasoning here because it uses algebraic examples to clarify what has been a complex issue.  First consider the possible behaviour of $G(Q^2)$ in the absence of QCD's scaling violations.  The transition form factor involves only one off-shell photon; and therefore a vector meson dominance (VMD) \emph{Ansatz} for the $Q^2$-dependence of the transition produces
\begin{equation}
\label{GVMDQ}
2 f_\pi G(Q^2)=m_\rho^2/(m_\rho^2 + Q^2)\,.
\end{equation}
(\emph{N.B}.\ The analogue for $F_\pi(Q^2)$ bounds the empirical data from above; but this is not the case with the transition form factor.)

\begin{figure}[t]
\centerline{\includegraphics[clip,width=0.6\textwidth]{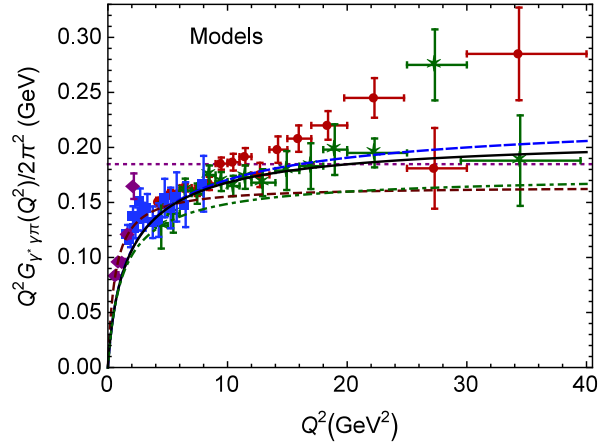}}

\caption{\label{ModelTFF}
$Q^2 G(Q^2)/(2\pi^2)$, model results.
Curves:
dotted (purple) -- asymptotic limit, derived from Eq.\,\eqref{BLuv};
dashed (brown) -- VMD result, derived from Eq.\,\eqref{GVMDQ};
dot-dashed (green) -- result obtained from Eq.\,\eqref{anomalytriangle} using a pion Bethe-Salpeter amplitude that generates $\varphi^{\rm cl}(x)$ in Eq.\,\eqref{PDAcl} using Eqs.\,\eqref{pionPDA}, \eqref{piWFA};
long-dashed (blue) -- result obtained from Eq.\,\eqref{anomalytriangle} using a pion PDA that generates $\varphi_\pi(x;\zeta_H)$ in Eq.\,\eqref{phiH} using Eqs.\,\eqref{pionPDA}, \eqref{piWFA};
solid (black) -- result obtained from Eq.\,\eqref{anomalytriangle} using a PDA that evolves from  $\varphi_\pi(x;\zeta_H)$ in Eq.\,\eqref{phiH} to $\varphi_\pi(x;\zeta=Q)$ in Eq.\,\eqref{alphanu} on $Q>\zeta_H$.
Data: BaBar \cite{Aubert:2009mc} -- circles (red); CELLO \cite{Behrend:1990sr} -- diamonds (purple); CLEO \cite{Gronberg:1997fj} -- squares (blue); Belle \cite{Uehara:2012ag} -- stars (green).}
\end{figure}

The expression in Eq.\,\eqref{GVMDQ} yields a $Q^2\approx 0$ slope for the transition form factor (a neutral pion radius, $r_{\pi^0}$) that is consistent with data \cite{Agashe:2014kda}: $r_{\pi^0} \approx r_{\pi^+}$, so this expression must be a reasonable approximation on $Q^2 \simeq 0$.  However, as evident in Fig.\,\ref{ModelTFF}, it approaches an asymptotic limit of $m_\rho^2/[2f_\pi]$, which is just 90\% of the result associated with Eq.\,\eqref{BLuv}. Consequently, if $G(Q^2)$ is to approach the limit in Eq.\,\eqref{BLuv} from above in the absence of scaling violations, then it must be influenced by at least three distinct mass scales:
an infrared scale that fixes the transition radius, which is smaller than the scale associated with the limit in Eq.\,\eqref{BLuv};
an intermediate scale, marking the point at which nonperturbative aspects of the pion's internal structure begin to take full control of the transition so that the function's fall-off rate may slow and $2f_\pi G(Q^2)$ can thereafter evolve to lie above the ultraviolet limit;
and finally an ultraviolet scale, at which the fall-off rate increases again, as required if $Q^2 G(Q^2)$ is to reach the limit in Eq.\,\eqref{BLuv}.
In the absence of scaling violations, the existence of three mass scales is unlikely, especially since just two scales are evident in predictions for $F_\pi(Q^2)$ described in Sec.\,\ref{SecElasticPion} and the active elements are identical: in both cases, one off-shell photon and the pion wave function.  If three mass scales were possible, they would most probably appear in $F_\pi(Q^2)$ because the associated process is influenced by two pion wave functions in contrast with the single wave function involved in $G(Q^2)$.

%
%
%
%
%

Scaling violations are, however, a feature of QCD.  Thus a third scale may appear in connection with the transition form factor, \emph{viz}.\ that associated with the progression to perturbative QCD, which is expressed in the appearance of an additional logarithmic momentum dependence, $[\ln Q^2/\Lambda_{\rm QCD}^2]^{{\mathpzc p}_G}$, that amends dimensional power-law behaviour in the ultraviolet.  The momentum scale for this progression in the charged-pion form factor is $Q^2 \approx 8\,$GeV$^2$ [Sec.\,\ref{SecElasticPion}].  Universality of pion structure in related processes suggests that a similar scale should be active in the transition form factor.  Considering the neighbourhood $Q^2\simeq 8\,$GeV$^2$, all empirical results for $Q^2 G(Q^2)$ lie below the limit in Eq.\,\eqref{BLuv}; and hence, if logarithmic evolution of $Q^2 G(Q^2)$ becomes established on this domain, then the ultraviolet limit would still be approached slowly from below.  However, if the progression domain is broadened in this case, owing to the presence of just one pion in the transition process, then $Q^2 G(Q^2)$ might grow to marginally exceed $2f_\pi$ before ${\mathpzc p}_G(Q^2)$ finally acquires that asymptotic value which describes a slow approach to the limit in Eq.\,\eqref{BLuv}.

It is this line of reasoning which leads to a picture of $G(Q^2)$ that involves three mass scales: one associated with the radius, blending effects from the photon-quark interaction and pion structure; another characterising the domain upon which nonperturbative features of pion structure fully control the $\gamma^\ast \gamma \to \pi^0$ transition; and a third typifying the region within which the magnitude of $G(Q^2)$ is still fixed by nonperturbative physics but the momentum dependence of $[Q^2 G(Q^2)-2f_\pi]$ has acquired $[\ln Q^2/\Lambda_{\rm QCD}^2]^{{\mathpzc p}_G}$-damping characteristic of scaling violations in QCD.  One should consequently expect that the limit in Eq.\,\eqref{BLuv} will either be approached from below or only exceeded slightly, perhaps on a broad domain, with logarithmic approach to the ultraviolet limit in either case.

A unification of the $\gamma_Q^\ast \gamma_Q^\ast \to \pi^0$, $\gamma_Q^\ast \gamma \to \pi^0$ transition and charged-pion elastic form factors on $0\leq Q^2 \leq 4\,$GeV$^2$ was accomplished in Ref.\,\cite{Maris:2002mz}.  In fact, the $\gamma_Q^\ast \gamma_Q^\ast \to \pi^0$ transition was computed to arbitrarily large $Q^2$ and shown to approach its QCD hard-photon limit uniformly from below.  However, as explained in connection with the elastic form factor in Sec.\,\ref{SecElasticPion}, using the simple algorithm employed in Ref.\,\cite{Maris:2002mz} it is impossible to extend the $\gamma_Q^\ast \gamma \to \pi^0$ transition calculations into the domain of momenta relevant to modern experiments.  Again, that difficulty is overcome using the methods introduced in Ref.\,\cite{Chang:2013pqS} for computation of the pion's leading-twist PDA, which not only enable the computation of $F_\pi(Q^2)$ to arbitrarily large-$Q^2$, as described in Sec.\,\ref{SecElasticPion}, but also $G(Q^2)$.

The $\gamma^\ast \gamma \to \pi^0$ transition form factor is computed from
${\mathpzc T}_{\mu\nu}(k_1,k_2) = T_{\mu\nu}(k_1,k_2)+ T_{\nu\mu}(k_2,k_1)$,
where the pion's momentum $P=k_1+k_2$, $k_1$ and $k_2$ are the photon momenta; and, at the same order of truncation used for $F_\pi(Q^2)$ in Eq.\,\eqref{RLFpi},
\begin{eqnarray}
\nonumber
T_{\mu\nu}(k_1,k_2) &= & \frac{e^2}{4\pi^2}\, \epsilon_{\mu\nu\alpha\beta}\,k_{1\alpha} k_{2\beta}\, G(k_1^2,k_1\cdot k_2,k_2^2)\\
& =& \! {\rm tr}\!\!\! \int\!\!\!\frac{d^4 \ell}{(2\pi)^4} \,
i {\cal Q} \chi_\mu(\ell,\ell_1)\, \Gamma_\pi(\ell_1,\ell_2) \, S(\ell_2) \, i {\cal Q} \Gamma_\nu(\ell_2,\ell)\,.
\label{anomalytriangle}
\end{eqnarray}
Here $\ell_{1}=\ell+k_1$, $\ell_{2} = \ell - k_2$, ${\cal Q} = {\rm diag}[e_u,e_d] = e\, {\rm diag}[2/3,-1/3]$, where $e$ is the positron charge, and $\Gamma_\nu$ is the amputated photon-quark vertex.  The kinematic constraints are $k_1^2=Q^2$, $k_2^2=0$, $2\, k_1\cdot k_2=- (m_\pi^2+Q^2)$; and the manner by which Eq.\,\eqref{anomalytriangle} provides for a parameter-free realisation of Eq.\,\eqref{Ggpg0} is detailed in Refs.\,\cite{Roberts:1994hh, Maris:1998hc, Holl:2005vu}.  A pictorial representation of Eq.\,\eqref{anomalytriangle} may be drawn similar to Fig.\,\ref{FpiVertex}, but with the rightmost pion Bethe-Salpeter amplitude replaced by the amputated photon-quark vertex.

The computation of $G(Q^2)$ in Ref.\,\cite{Raya:2015gvaS} was deliberately formulated so as to deliver a coherent picture of both the transition and elastic form factor, reviewed in Sec.\,\ref{SecElasticPion}, and therefore used forms for all elements common to Eqs.\,\eqref{RLFpi} and \eqref{anomalytriangle} that are consistent with those in Ref.\,\cite{Chang:2013niaS}.  Notably, whilst there is the appearance of a difference between the representations of the photon-quark vertex in these two studies, they are effectively equivalent.  With all elements in Eq.\,\eqref{anomalytriangle} thus determined, computation of $G(Q^2)$ is straightforward.  Before proceeding to explain the complete result, however, Ref.\,\cite{Raya:2015gvaS} used simple examples and reasoning to elucidate some essential features of the transition form factor; a discussion which we summarise below.

Evolution of the pion's PDA with the resolving scale $\zeta$ is logarithmic \cite{Lepage:1979zb, Efremov:1979qk, Lepage:1980fj}; and, as we indicated in Sec.\,\ref{SecElastic}, whilst Poincar\'e covariant computations using a renormalisation-group-improved RL truncation produce the same matrix-element power-laws as perturbative QCD, they fail to reproduce the full anomalous dimensions \cite{Lepage:1980fj}.  Typically \cite{Maris:1998hc, Chang:2013pqS, Chang:2013niaS}, the RL approximation to a matrix element underestimates the rate of its logarithmic flow with the active momentum scale because the approximation omits gluon-splitting diagrams.  As observed in Ref.\,\cite{Raya:2015gvaS}, owing to Eq.\,\eqref{pionPDA}, the pion's Poincar\'e covariant Bethe-Salpeter wave function must evolve with $\zeta$ in the same way as $\varphi_\pi$, but this constraint had generally been overlooked in computations of observables using continuum methods in QCD.  Such evolution is important: it enables the dressed-quark and -antiquark degrees-of-freedom, in terms of which the wave function is expressed at a given scale $\zeta^2=Q^2$, to split into less well-dressed partons via the addition of gluons and sea quarks in the manner prescribed by QCD dynamics.  These and similar processes are incorporated naturally in bound-state problems when the complete quark-antiquark scattering kernel is used; but aspects are lost when that kernel is truncated, and so it is with the RL truncation.

The impact of this realisation on $G(Q^2)$ may be illustrated by considering the Bethe-Salpeter wave function constructed from
\begin{subequations}
\label{piWFA}
\begin{eqnarray}
S(k) & =& 1/[i\gamma\cdot k + M] \,,\\
\label{pionBSAmodel}
{\mathpzc n}_\pi\, \Gamma_\pi(k;P) & =& i\gamma_5\,\frac{M}{f_\pi}\,\int_{-1}^1\! dz\,\rho_\nu(z)\frac{M^2}{(k + z P/2)^2+\Lambda_\pi^2}\,,\\
\rho_\nu(z) & =& \frac{\Gamma(\frac{3}{2}+\nu)}{\sqrt{\pi}\,\Gamma(1+\nu)}\,(1-z^2)^\nu\,. \label{rhonu}
\end{eqnarray}
\end{subequations}
Inserting these formulas into Eq.\,\eqref{pionPDA}, then with $\nu=1$ the result is $\varphi^{\rm cl}(x)$ in Eq.\,\eqref{PDAcl}, \emph{i.e}.\ the PDA associated with QCD's conformal limit.  Hence, a computation of the transition form factor in Eq.\,\eqref{anomalytriangle} using Eqs.\,\eqref{piWFA} with $\nu=1$ will yield a result definitive of a conformal-limit pion Bethe-Salpeter wave function, \emph{viz}.\ a wave function that is frozen to produce Eq.\,\eqref{PDAcl} at all scales $\zeta$.  Such a calculation yields the dot-dashed (green) curve in Fig.\,\ref{ModelTFF}, which is a monotonically increasing, concave function that approaches the asymptotic limit associated with Eq.\,\eqref{BLuv} from below.

As emphasised in connection with Eq.\,\eqref{phiH}, however, the PDA at any scale realisable with contemporary facilities is very different from $\varphi^{\rm cl}(x)$.  The concave, dilated PDA in Eq.\,\eqref{phiH} is obtained from Eq.\,\eqref{pionPDA} by using \mbox{$\nu=-1/2$} in Eqs.\,\eqref{piWFA}.   This knowledge enables one to compute $G(Q^2)$ via Eq.\,\eqref{anomalytriangle} using a Bethe-Salpeter wave function that is frozen to produce Eq.\,\eqref{phiH} at all scales $\zeta$.  The result is the long-dashed (blue) curve in Fig.\,\ref{ModelTFF}.  Like the prediction obtained using $\nu=1$, this curve is monotonically increasing and concave, and approaches its asymptotic limit from below.  The difference is that the asymptotic limit is not that associated with Eq.\,\eqref{BLuv}.  Instead, this curve approaches $(8/3) f_\pi$ as $Q^2\to \infty$.

A last illustration enables a unifying thread to be drawn between the material detailed above and analyses of the neutral pion transition form factor that have attempted to reconcile the BaBar measurements with Eq.\,\eqref{BLuv}, \emph{e.g}.\ Refs.\,\cite{Radyushkin:2009zg, Agaev:2010aq, Dorokhov:2013xpa}.  Consider therefore $2\rho_\nu(z)=\delta(1+z)+\delta(1-z)$ in Eq.\,\eqref{pionBSAmodel}.  This gives $\varphi_\pi(x)=1$ via Eq.\,\eqref{pionPDA}, which is also the result obtained using a translationally invariant regularisation of a momentum-independent quark-quark interaction \cite{Roberts:2010rnS}.  (Some treatments of the Nambu--Jona-Lasinio model fall in this class.)  With this input, Eq.\,\eqref{anomalytriangle} yields $Q^2 G(Q^2)  \propto [\ln Q^2/M^2]^2$, and the mass-scale $M$  can be tuned to reproduce the BaBar data in Fig.\,\ref{ModelTFF}.  Notwithstanding that, the complete curve is also monotonically increasing and concave, and approaches its asymptotic limit from below \cite{Roberts:2010rnS}.  Notably, the BaBar data have often been used to justify a ``flat-top'' pion PDA: $\varphi_\pi(x)\approx 1$.  Employing the factorised hard-scattering formula in this case, one finds $Q^2 G(Q^2)  \propto [\ln Q^2/M^2]$, a result which highlights an observation made above, \emph{viz}.\ Poincar\'e covariant treatments of the triangle diagram expressed by Eq.\,\eqref{anomalytriangle} typically yield the correct power-law but produce an erroneous value of the anomalous dimension.

Evidently, any computation of $G(Q^2)$ via Eq.\,\eqref{anomalytriangle} which uses a Bethe-Salpeter amplitude that does not evolve with the resolution scale, $\zeta^2=Q^2$, produces a curve $Q^2 G(Q^2)$ which approaches its asymptotic limit from below; but the value of that limit depends on the model used for the pion's (frozen) Bethe-Salpeter wave function \cite{Roberts:2010rnS}.

Evolution of the interaction current and vertices in Eq.\,\eqref{anomalytriangle} is missing from such model calculations \cite{Raya:2015gvaS}; but in the context of Eqs.\,\eqref{pionPDA}, \eqref{piWFA}, it can be translated into an evolution of $\nu$ in Eq.\,\eqref{rhonu}, \emph{i.e}.\ one can reproduce any concave PDA $\varphi_\pi(x;\zeta)$,  obtained via ERBL evolution of $\varphi_\pi(x;\zeta_H)$, by using a suitably chosen value of $\nu(\zeta)$.
Solving for $\nu(\zeta)$ is straightforward because at any $\zeta>\zeta_H$, the ERBL-evolved form of $\varphi_\pi(x;\zeta_H)$ is
\begin{equation}
\varphi_\pi(x;\zeta) = [x(1-x)]^{\alpha(\zeta)}/B(1+\alpha(\zeta),1+\alpha(\zeta))\,.
\label{alphanu}
\end{equation}
On the domain $\alpha(\zeta) \in (0.3,0.8)$ that is relevant to contemporary experiment and theory \cite{Raya:2015gvaS}: \begin{equation}
\label{nuofalpha}
\nu(\alpha(\zeta)) = - [ 5.4 - 6.6\,\alpha(\zeta) ]/[ 5.9-2.8 \,\alpha(\zeta)]\,.
\end{equation}

Employing Eqs.\,\eqref{piWFA}, \eqref{nuofalpha} in Eq.\,\eqref{anomalytriangle}, one obtains a result for $G(Q^2)$ that expresses the impact of a Bethe-Salpeter wave function which evolves and thereby interpolates between $\varphi_\pi(x;\zeta_H)$ in Eq.\,\eqref{phiH} and $\varphi^{\rm cl}(x)$ in Eq.\,\eqref{PDAcl}.  This is the solid (black) curve in Fig.\,\ref{ModelTFF}.  Obtained this way, $Q^2 G(Q^2)$ is monotonically increasing and concave on the domain depicted, and reaches a little above the asymptotic limit associated with Eq.\,\eqref{BLuv}.  The growth is logarithmically slow, however; and whilst the curve remains a line-width above the asymptotic limit on a large domain, logarithmic growth eventually becomes suppression, and the curve thereafter proceeds towards the QCD asymptotic limit from above.  One thus has a simple illustration and concrete realisation of the picture drawn in the penultimate paragraph preceding Eq.\,\eqref{anomalytriangle}.

These remarks acquire additional meaning within the context provided by the leading-twist expression for the transition form factor \cite{Lepage:1980fj}:
\begin{equation}
\label{HardG}
G(Q^2) = 4\pi^2f_\pi\int_0^1 \! dx \,T_H(x,Q^2,\alpha(\zeta);\zeta)\,\varphi_\pi(x;\zeta)\,,
\end{equation}
where $T_H(\zeta)$ is the photon$+$quark$+$antiquark scattering amplitude appropriate to the scale $\zeta$.  On the domain $\Lambda_{\rm QCD}/\zeta \simeq 0$, $T_H(\zeta) = (e_u^2-e_d^2)/(xQ^2)$.  However, this is far from an accurate representation of the scattering amplitude at an hadronic scale, $\zeta=\zeta_H \approx 2\,$GeV, a fact made plain by Eq.\,\eqref{anomalytriangle}, which involves nonperturbatively dressed propagators and vertices.  Indeed, from a light-front perspective, this dressing corresponds to inclusion of infinitely many Fock-space components in the description of the pion bound-state, its constituents and their interactions.  Now, owing to convergence issues connected with the need to extrapolate from $\Lambda_{\rm QCD}/\zeta \simeq 0 \to \Lambda_{\rm QCD}/\zeta_H$, a twist expansion cannot systematically connect Eq.\,\eqref{HardG} with Eq.\,\eqref{anomalytriangle}.  On the other hand, if RL-truncation DSE solutions are used for the dressed propagators and vertices in Eq.\,\eqref{anomalytriangle}, then one arrives at Eq.\,\eqref{HardG} on $\Lambda_{\rm QCD}/\zeta \simeq 0$, except for a mismatch $\sim [\ln Q^2/\Lambda_{\rm QCD}^2]^{{\mathpzc p}_G}$.  As we have explained, this discrepancy originates in the failure of RL truncation to reproduce the complete array of gluon and quark splitting effects contained in QCD and hence its failure to fully express interferences between the anomalous dimensions of those $n$-point Schwinger functions which are relevant in the computation of a given scattering amplitude.  It is ameliorated by the procedure discussed in connection with Eq.\,\eqref{alphanu}.  Naturally, if a similar procedure is employed in revisiting the kindred calculation of the pion form factor, described in Sec.\,\ref{SecElasticPion}, one can also correct the $\ln Q^2$-evolution of the RL prediction for $F_\pi(Q^2)$.

%
%
%
%
%

\begin{figure}[t]
\centerline{\includegraphics[clip,width=0.6\textwidth]{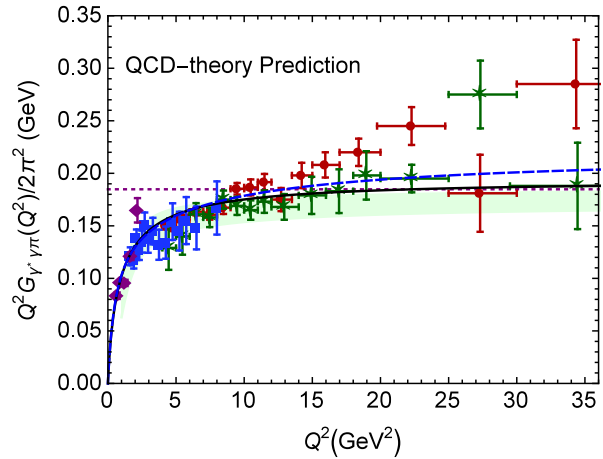}}

\caption{\label{figDSEprediction}
Prediction for $Q^2 G(Q^2)/(2\pi^2)$ .
Curves:
solid (black) -- result obtained from Eq.\,\eqref{anomalytriangle} using RL-truncation propagators, amplitudes and vertices, and ERBL evolution of the pion Bethe-Salpeter amplitude;
long-dashed (blue) -- result obtained without that evolution;
dotted (purple) -- asymptotic limit, derived from Eq.\,\eqref{BLuv}.
Data: BaBar \cite{Aubert:2009mc} -- circles (red); CELLO \cite{Behrend:1990sr} -- diamonds (purple); CLEO \cite{Gronberg:1997fj} -- squares (blue); Belle \cite{Uehara:2012ag} -- stars (green).
The shaded (green) band is described in Ref.\,\cite{Bakulev:2012nh}.
}
\end{figure}

The $\gamma^\ast \gamma\to \pi^0$ transition form factor calculated from Eq.\,\eqref{anomalytriangle}, using propagators and vertices that are consistent with those employed in computing $F_\pi(Q^2)$, is depicted as the dashed (blue) curve in Fig.\,\ref{figDSEprediction}.  The solid (black) curve in this figure is obtained from the dashed curve via multiplication by a $Q^2$-dependent evolution factor.  That factor is a ratio: the denominator is $G(Q^2)$ computed using Eqs.\,\eqref{anomalytriangle}, \eqref{piWFA} with a frozen value of $\alpha=0.3$, representing the RL result in Eq.\,\eqref{phiRLreal}; and the numerator is $G(Q^2)$ calculated using those same equations but with $\nu$ determined via Eq.\,\eqref{nuofalpha} as $\alpha$ experiences one-loop ERBL evolution on $Q^2>\zeta_H^2$.  This solid curve expresses what Ref.\,\cite{Raya:2015gvaS} judges to be its best prediction for the transition form factor.  The two salient features of the prediction are clear.  Notwithstanding the fact that $G(Q^2)$ was evaluated using a framework which produces a pion PDA that is a broad, concave function at the hadronic scale $\zeta_H$: the calculated transition form factor does not materially exceed the asymptotic limit in Eq.\,\eqref{BLuv}; and this same approach explains both existing measurements of $F_\pi(Q^2)$ and its hard-photon limit.  Importantly, the solid curve in Fig.\,\ref{figDSEprediction} behaves in precisely the manner one would expect based on the analyses of simple models described in connection with Fig.\,\ref{ModelTFF}.  Hence, one may argue that the marked similarity between the solid curves in Figs.\,\ref{ModelTFF} and \ref{figDSEprediction} means that the simple model expressed by Eqs.\,\eqref{piWFA}, \eqref{nuofalpha} contains all that is essential to fully comprehend the nature of the transition form factor.

In the context of a survey of theoretical analyses of the $\gamma^\ast \gamma \to \pi^0$ transition \cite{Bakulev:2012nh}, the prediction in Ref.\,\cite{Raya:2015gvaS} is a member of that class of studies, denoted by the (green) shaded band  which are consistent with all non-BaBar data and confirm the standard QCD factorisation result in Eq.\,\eqref{BLuv}.  In particular, the solid (black) curve in Fig.\,\ref{figDSEprediction} is similar to the light-cone sum rules result of Ref.\,\cite{Bakulev:2011rp} on their common domain: sum rules analyses are restricted to  $Q^2 \gtrsim 1\,$GeV$^2$.  It is worth remarking, too, that on $Q^2\gtrsim 10\,$GeV$^2$, the solid (black) curve in Fig.\,\ref{figDSEprediction} also matches the AdS/QCD model result in Ref.\,\cite{Brodsky:2011xx}.

One may contrast the scheme for calculating $G(Q^2)$ described above, which connects a prediction for the pion's PDA with the $Q^2$-dependence of the transition form factor, with the class of approaches that choose instead to infer a form of $\varphi_\pi$ by requiring agreement with the BaBar data, \emph{e.g}.\ Ref.\,\cite{Agaev:2010aq}.  A point of comparison is provided by Table~II therein, which lists Gegenbauer-$3/2$ moments associated with the PDAs judged viable by this criterion.  These moments are defined via
\begin{equation}
\label{projection}
a_j(\zeta) = \frac{2}{3}\ \frac{2\,j+3}{(j+2)\,(j+1)}\int_0^1 dx\, C_j^{(3/2)}(2\,u-1)\,\varphi_\pi(u;\zeta)\,,
\end{equation}
where $\{C_j^{(3/2)},j=1,\ldots,\infty\}$ are Gegenbauer polynomials of order $\alpha=3/2$.  Using the PDAs in Eqs.\,\eqref{phiH}, \eqref{phiRLreal} one finds (at $\zeta_H=2\,$GeV):
\begin{equation}
\label{Gmoments}
\begin{array}{lcccccc}
    & a_2 & a_4 & a_6 & a_8 & a_{10}& a_{12}\\
{\rm DB} & 0.15 & 0.057 & 0.031 & 0.018 & 0.013 & 0.0093 \\
{\rm RL} & 0.23 & 0.11\phantom{7} & 0.066 & 0.045 & 0.033 & 0.025\phantom{3}
\end{array}\,.
\end{equation}
The root-mean-square relative-difference between these moments and those determined in Ref.\,\cite{Agaev:2010aq} is roughly 100\%.  Thus the PDA needed to reproduce the BaBar data is irreconcilable with that determined in an \emph{ab initio} computation that unifies the electromagnetic form factors of the charged and neutral pions.

The computation of the $\gamma^\ast \gamma \to \pi^0$ transition form factor \cite{Raya:2015gvaS} reviewed here completes a unified description and explanation of this transition with the charged pion electromagnetic form factor, its valence-quark distribution amplitude, and numerous other quantities \cite{Roberts:2000aa, Chang:2011vu, Bashir:2012fs, Cloet:2013jya}.  Importantly, it demonstrates that a fully self-contained and consistent treatment can readily connect a pion PDA that is a broad, concave function at the hadronic scale with the perturbative QCD prediction for the transition form factor in the hard photon limit.  As evident in Fig.\,\ref{figDSEprediction}, the prediction for $G(Q^2)$ agrees with all available data, except that obtained by the BaBar collaboration, and is fully consistent with the hard scattering limit.  It is worth reiterating that the normalisation of the $\gamma^\ast \gamma \to \pi^0$  transition form factor's hard scattering limit is set by the pion's leptonic decay constant, whose magnitude is fixed by the scale of dynamical chiral symmetry breaking, a crucial feature of the Standard Model.  Therefore, in order to claim an understanding of the Standard Model, it seems critical to obtain new, accurate and precise transition form factor data on $Q^2>10\,$GeV$^2$ so that contemporary predictions can reliably be tested.

\section{Epilogue}
Our knowledge of the nature of the pion and its properties has evolved continuously since its discovery almost seventy years ago; but, during the last fifteen years, our ability to probe the pion's interior and develop a clear picture of its internal structure has received a great boost from experiments enabled by modern facilities.  Almost in tandem, there have been major advances in treating the continuum bound-state problem in QCD so that new predictions have been made, which serve to motivate further experiments at upgraded facilities.  These new studies can potentially provide conclusive answers to some longstanding questions and supply evidence needed to incorporate the kaon into a gestalten whole.  The road to complete understanding is long.  It has been and continues to be travelled by many wanderers; and it is too early to say the end is in sight.  Notwithstanding that, the journey has revealed a great deal about the origin of visible mass in the Universe.  The next steps must be to consolidate the picture that has emerged; and to detail the role played by the pion, and Nambu-Goldstone modes generally, in the confinement of systems constituted from quarks and gluons.

Confinement is a key problem in modern physics and its solution is unlikely to be found through theoretical analysis alone.  A multipronged approach is required, involving constructive feedback between experiment and theory in studies of, \emph{e.g}.: the production and spectrum of excited and exotic hadrons; hadron elastic and transition form factors; generalised and transverse momentum dependent parton distributions, and particularly fragmentation functions.  Experimental data, and its correlation and parametrisation using models will be essential in this interrelated effort.  However, as the discussion herein highlights, there is an overriding need for reliable, QCD-connected calculations of all these quantities, for only following the comparisons that such information enables can we truly begin to arrive at an understanding of QCD, the strong interaction sector of the Standard Model.

\ack

Both the results described and the insights drawn herein are fruits from collaborations we have joined with many colleagues and friends throughout the world; and we are very grateful to them all.
%
This work was supported in part by NSF grant PHY-1306227;
and by the U.S.\ Department of Energy, Office of Science, Office of Nuclear Physics, under contract no.~DE-AC02-06CH11357.

\newpage


\providecommand{\newblock}{}

\newpage

\tableofcontents

\end{document}